\documentclass[11pt,titlepage]{article}
\pdfoutput=1
\usepackage[margin=1in]{geometry}
\usepackage{xcolor}
\usepackage{hyperref}
\usepackage{amsmath}
\usepackage{array}
\usepackage{graphicx}
\usepackage{subcaption}
\usepackage{authblk}

\usepackage[autostyle=false, style=english]{csquotes}
\MakeOuterQuote{"}

\renewcommand\footnotemark{}

\title{Small World Student Network at the University of Texas at Dallas in Times of Social Distancing
\thanks{The authors wish to thank Christopher Simmons and Jaynal Pervez of the Office of Information Technology at UT Dallas for their help in using the high performance computing cluster for data analysis.}
}
\author{Kailash Subramanian}
\author{Joshua M. Williams}
\author{Daniel C. DeAnda}
\author{Aditya A. Agrawal}
\author{Andrei Racila}
\author{Aditi R. Prabhu}
\author{Lawrence Redlinger}
\author{Christopher Wendt}
\author{Ravi Prakash}
\affil{University of Texas at Dallas, Richardson, Texas 75080, U.S.A.}

\begin{document}

\maketitle

\thispagestyle{empty}

\begin{abstract}
To limit the spread of the novel coronavirus on college campuses, a common strategy for the Fall 2020 and Spring 2021 terms has been to offer instruction weighted toward hybrid or fully online modalities. Colleges are now considering whether and how to expand hybrid or fully in-person instruction for future terms, and learn lessons from this experience for future use. Our paper uses Fall 2019 enrollment data for a medium-sized public American university to analyze whether some student groupings by class standing or course level are more susceptible to the spread of infectious disease through academic enrollment networks. Replicating Weeden and Cornwell \cite{Weeden_Cornwell_2020}, we find that enrollment networks at the institution are "small worlds" characterized by high clustering, short average path lengths, and multiple independent connections. We also find that connectivity decreases as class standing (graduate vs. undergraduate; senior vs. freshman) and course level increase; as students move from generalized to specialized course loads, networks cluster by major. Holding other factors constant, our analysis indicates that policies focusing on in-person instruction for lower division students to capture the freshman experience conflict with the greater risk of infectious spread through a lower division network in the absence of additional steps to minimize academic connectivity. There are academic and financial incentives for emphasizing the freshman experience, including concerns about student attrition from the first to second academic year and recouping costs of infrastructure investments in dormitories that are disproportionately used by freshmen. Possible solutions could include (i) restricting face-to-face or hybrid instruction to courses in students' academic majors, which would disrupt larger networks into smaller ones and thus restrict the spread of infection across majors, and (ii) take a "scalpel" approach to instruction modes by moving online courses most likely to facilitate epidemic spread.
\end{abstract}

\section{Introduction}
\label{Introduction}

The COVID-19 epidemic resulted in the suspension of face-to-face teaching on university campuses in the United States in the spring of 2020. The subsequent debate on when and how to return to face-to-face instruction has been informed by several recent studies. Notably, Weeden and  Cornwell \cite{Weeden_Cornwell_2020} studied course enrollment at Cornell University and concluded that a graph with edges between students registered for the same course section forms a small world network, with high clustering, short average path length, and multiple independent paths between pairs of students. This appears to indicate that infections could spread quickly among the student population unless several courses were converted to online instruction, with face-to-face instruction limited to small sections only. 

Given that a vaccine is unlikely to produce community immunity before the start of the fall semester of  2021 and fully online instruction is considered suboptimal for student learning and institutional finances, the key question for the remainder of Academic Year 2021 is to what extent institutions can return to face-to-face or hybrid instruction. How many classes can resume face-to-face or hybrid instruction, how large can they be, and which students can populate them? Also, what are the lessons to be learned from this experience so that we are better prepared for course planning and scheduling if such epidemics are in the our future?


This paper describes an effort to replicate Weeden and Cornwell's study in the context of the University of Texas at Dallas (henceforth, "UT Dallas"), a medium-sized public university. Cornell and UT Dallas have a similar enrollment, though UT Dallas has a higher share of undergraduates ($71\%$ vs. $63\%$ per the 2019-2020 Common Data Set), a higher ratio of incoming transfer students to incoming freshmen ($1:2$ vs. $1:6$), and an undergraduate population that is older on average ($13\%$ of UT Dallas undergraduates are over $25$, vs. $1\%$ at Cornell University). UT Dallas has a smaller fraternity/sorority population ($\sim 5\%$, vs. $\sim 25\%$) and a smaller share of freshmen ($52\%$, vs. $99\%$) and undergraduates ($24\%$, vs. $52\%$) living on campus. In addition, UT Dallas has a higher student/instructor ratio ($24:1$ vs. $9:1$). Whereas $56\%$ of Cornell’s undergraduate sections have fewer than $20$ students, at UT Dallas the share is only $20\%$. Most critically for our study, UT Dallas has a core curriculum requirement for all undergraduate students, under which students must take some courses outside their discipline (usually as freshmen or sophomores). Subsequent sections of this paper discuss whether these differences result in any significant impact on clustering, average path length, and bi-connectivity of students.


The Weeden and Cornwell study analyzed enrollment data for the entire university, for undergraduate students only, and for the Liberal Arts College. For the UT Dallas study, we perform a similar analysis for the entire university, as well as separately for graduate and undergraduate students. Additionally, due to the general core requirement (which for the most part correspond to lower division courses, i.e., those at the freshmen and sophomore levels), we separately analyze the data for lower division and upper division undergraduate courses. We also investigate the relationships at inter-school level. 

While a $1000$-level course at UT Dallas is usually interpreted as a freshman-level course, the reality is a little different. Even though most $1000$-level courses primarily have freshmen enrolled in them, some also include sophomores, juniors and seniors trying to fulfill general core requirements. Likewise, freshmen with ample AP and other college-level credits may be taking higher level courses. Therefore, we also analyze networks formed among undergraduate students by considering the graphs formed using only (i) edges between students taking courses at the same ($1000, ~2000, 3000, or~ 4000$) level, and (ii) edges between students at the same class rank, namely freshmen, sophomores, juniors, or seniors. Several interesting trends emerge when juxtaposing these networks together. 

As described in Section~\ref{Analysis_Fall19}, we notice a pattern that calls into question the efficacy of the strategy followed by some universities as they started classes in Fall 2020. We observed that all students taking freshmen-level classes form a highly connected and low-diameter network, with students across all majors reachable from each other along short paths. By comparison, graduate students - and to a slightly lesser extent those taking senior-level classes - tend to exhibit a higher level of clustering by major or school. Students taking graduate and senior level courses from the same school are densely clustered together with several short paths among them, and there are significantly fewer links connecting these intra-school or intra-major clusters. \textbf{This leads us to conclude that, rather than inviting freshmen to take in-person classes, it would have been more prudent to limit in-person classes to graduate students and upper-division students while offering only online/remote instruction for lower-division courses.}

\section{Related Work}
\label{Related}

Milgram and Travers  introduced the notion of small-world networks among people \cite{milgram1967small,travers1977experimental}. In \cite{travers1977experimental}, they showed that participants in Nebraska and Boston managed to get letters delivered to targets in Massachusetts using relatively short chains of acquaintances. They also refer to the presence of a small number of "sociometric stars" through which a disproportionate number of acquaintance chains passed. Their work spurred investigation of small-world phenomenon in a number of disciplines, including sociology, medicine, and computer science. Graph theory has been applied to this phenomenon by all of these disciplines. People are modeled as nodes in a graph. An edge is drawn between two nodes if a relation relevant to the problem being studied exists between the people corresponding to those nodes.

Watts and Strogatz \cite{watts1998collective} proposed a framework for modeling networks ranging from truly random networks with low diameters to uniform networks with high diameters. They considered a uniform lattice with edges connecting neighboring nodes and only a small number of edges connecting neighbors replaced by edges between random pairs of nodes. As the fraction of these random edges increases, the graph transitions from uniform to random. Watts and Strogatz showed that, even with a small fraction of edges between distant nodes, the diameter of the graph reduces significantly and the graph starts exhibiting the small-world phenomenon. One effect of this phenomenon is that the diameter of the graph becomes a logarithmic function of the number of vertices. This is relevant to our study of student enrollment in courses. A two-mode graph of course enrollment would have students and courses as nodes, with an edge between student $x$ and course $y$ if student $x$ is enrolled in course $y$.  From such a graph, it is easy to obtain a single-mode graph with only students as nodes and edges between pairs of students taking the same course. While students of the same rank (freshmen, sophomore, etc.) and the same major may take most of their courses with other students of the same rank and major, there are always a small number of edges connecting \textit{random} students. For example, a computer science freshman may be in the same statistics class as a junior majoring in sociology.

Amaral et al. [1] studied three classes of small-world networks, namely scale-free networks, broad-scale networks, and single-scale networks. Scale-free networks in particular are small-world networks in which new vertices will preferentially connect to the more highly connected vertices within the network, wherein the diameter will increase logarithmically with the number of vertices. As such, the growth of a scale-free network closely mimics the transmission of an infectious disease through a population, where a person is more likely to be infected through a connection with a \textit{superspreader} who has contact with many people. 

The "sociometric stars" observed by Travers and Milgram in their experiments \cite{travers1977experimental} point to an important concept in graphs: \textit{point centrality}. Freeman \cite{freeman1977set,freeman1978centrality} described this as the \textit{betweenness} property of a vertex. A vertex has high centrality if it falls on the shortest path between a large number of pairs of vertices. In effect, this vertex can connect many otherwise distant pairs or clusters of vertices. From a course enrollment perspective, consider a mathematics major at UT Dallas. Though the number of students majoring in mathematics is small, this particular mathematics major may be in an algorithms class with computer science majors, a statistics course with business administration majors, and a calculus course with physics and mechanical engineering majors. This student would connect otherwise disparate clusters of students, lie on the shortest path between several pairs of students, and exhibit high centrality or betweenness. From an epidemiological perspective, if such a student were to contract an infectious disease like COVID-19, they could be a \textit{superspreader}. Hence, in our analysis of enrollment data, we attempt to identify the characteristics of students exhibiting high centrality.

Caley et al. \cite{caley2008quantifying} quantified the effect of social distancing on the spread of influenza in Sydney during the 1918-19 epidemic, as well as how successive waves of infection could be explained by varying levels of social distancing as people's perception of risk fluctuated. One way to minimize the risk of spreading an infectious disease through classroom attendance would be to move courses and students with high centrality to online instruction. This would have the effect of increasing network diameter and, therefore, enrollment-driven social distance between distinct groups of students.

As stated earlier, the work that is most relevant to the present study is by Weeden and Cornwell \cite{Weeden_Cornwell_2020}. They examine the potential for epidemic spread of an aerosolized virus (such as SARS-CoV-2) through in-person instruction at a medium-sized private American university (Cornell). Modeling connectivity between students co-enrolled in course sections, they find that the potential for spread is high due to college campuses meeting the criteria for a "small world" network \cite{watts1998collective}, namely high clustering (nearly all students are connected) and short average path lengths (the average student-pair has a low degree of separation). Despite academic concentration creating disciplinary silos, students are connected through hub courses such as general education requirements or gateway courses shared across disciplines. Furthermore, students are connected through multiple paths (shared sections or third parties), so removing even the most centralized sections or students will not significantly impact the diameter of the network. Weeden and Cornwell's analysis indicates that about half of all students were still connected in four or fewer steps even when all sections of 30 or more were eliminated from the network. Given full-time course loads, and that individuals can transmit SARS-CoV-2 for days before recognizing the need to quarantine, four steps is a good approximation of a week's worth of potential exposure on a college campus. Whereas Cornell's small class sizes means that a 30-student threshold forced the removal of approximately 16 percent of sections with an in-person component, at UT Dallas (a medium-sized, public university), a 30-student threshold affected $34\%$ of sections with an in-person component in Fall 2019, as well as $73\%$ of in-person student-section enrollments.

\section{UT Dallas Enrollment Data}
\label{Data}

For our primary analysis we consider student enrollment data for the University of Texas at Dallas, a medium-sized public university that is part of the University of Texas System. Total enrollment in the Fall 2019 semester was $29,449$ students registered in $5,235$ courses. As in the Weeden-Cornwell work, we removed all online courses from consideration and those students who registered only for such courses. This left us with $28,852$ students registered in $4,994$ courses. Of these,  $20,953$ were undergraduate students and $7,899$ were graduate students. Freshmen, sophomore, junior and senior level courses are numbered $1xxx$, $2xxx$, $3xxx$ and $4xxx$, respectively, while graduate level courses have higher numbers: $5xxx$ through $8xxx$. The second digit in each course number indicates the number of student contact hours per week for that course. Each course number is preceded by an alphabetic string representing the academic program that offers the course. Based on this prefix it is possible to uniquely associate a course with the school offering it.

The university is organized into eight schools, namely (i) Arts and Humanities (AH), (ii) Arts, Technology and Emerging Communication (ATEC), (iii) Behavioral and Brain Sciences (BBS), (iv) Economic, Political and Policy Sciences (EPPS), (v) Engineering and Computer Science (ECS), (vi) Interdisciplinary Studies (IS), (vii) Management (SOM), and (viii) Natural Sciences and Mathematics (NSM). Of these schools, Engineering and Computer Science and Management have the largest enrollment at both the undergraduate and graduate levels.

\subsection{Common Core}
\label{Common_core}

As our study is based on the one described in \cite{Weeden_Cornwell_2020}, it is important to state some of the important differences between Cornell and UT Dallas curriculum. While the exact number varies from one major to another, all undergraduate students need to complete just over 120 semester credit hours of coursework to graduate. Unlike Cornell, all undergraduate students at UT Dallas are required to complete a general education core curriculum of \textit{forty-two} semester credit hours, in addition to the required courses in their major, and any elective courses. 

The credit requirements for the general education core curriculum include six credit hours in Communication, three credit hours in Mathematics, six credit hours in Life and Physical Sciences, three credit hours in Language, Philosophy and Culture, three credit hours in Creative Arts, six credit hours in American History, six credit hours in Government/Political Science, three credit hours in Social and Behavioral Sciences, and six credit hours classified as Component Area Options. In each of these nine categories there are a large number of options. So, within the framework of core curriculum there is still a fair amount of flexibility. However, some courses like calculus, general chemistry, modern biology tend to have very high enrollment due to a large number of students interested in majoring in engineering, physical and life sciences related fields. These courses, for the most part, tend to have lectures with very large number of students, and laboratory sessions with comparatively smaller groups of students.

UT Dallas admits approximately $2,000$ transfer students every year who come in primarily as sophomores and juniors, having completed most of their general education core curriculum and some major-specific introductory courses at regional community colleges. Also, some UT Dallas students choose to take these general education core curriculum courses at community colleges and use the credits towards their graduation requirement. Courses taken at community colleges are not considered in our analysis. So, it is possible that based solely on UT Dallas courses we may infer that two students do not have any course in common when they may be taking a common course in a community college.


\subsection{Adjacency Matrix of Student Interaction}
\label{Adjacency_matrix}

We represent student enrollment information as a two-mode $n\times m$ matrix, $D$, where $n$ is the number of students, and $m$ is the number of sections offered across all courses (high enrollment courses do have multiple sections, with a subset of students taking the course getting enrolled in a given section). Students enrolled in the same section share the same physical space, while those in different sections do not. Each student is assigned a unique index in the range $[1,n]$. Similarly, each section is assigned a unique index in the range $[1,m]$. If student $i$ is enrolled in section $j$ then $D[i,j]=1$, otherwise $0$. From this matrix, we compute another matrix $A=D \times D^T$, where $D^T$ represents the transpose of matrix $D$. Matrix $A$ is an $n\times n$ matrix where:
\begin{equation}
\forall~ 1\leq i,j \leq n: A[i,j] = \sum_{k=1}^m D[i,k]*D^T[k,j]
\end{equation}
If students $i$ and $j$ are both enrolled in section $k$, then $D[i,k]*D^T[k,j]$ is equal to $1$. $A[i,j]$, where $i\neq j$ represents the number of sections in which students $i$ and $j$ are both enrolled. Hence, $A$ is a one-mode matrix that represents the strength of interaction between students: higher the value of $A[i,j]$, the greater the number of sections in which they are both enrolled and, therefore, the stronger the interaction between them. If two students, $x$ and $y$ are enrolled in no common section then $A[x,y]=0$. 

We also employ two variants of the $A$ matrix. We refer to them as $A_{binary}$ and $A_{weighted}$. Matrix $A_{binary}$ is a binary matrix such that if $A[i,j]>0$ then $A_{binary}[i,j]=1$, and if $A[i,j]=0$ then $A_{binary}[i,j]=0$. This is the matrix and its corresponding graph, described later in Section~\ref{Metrics}, that we use to compute the distance between students, average student network diameter, etc. 

The matrix $A_{weighted}$ attempts to capture the strength of interaction between pairs of students by not only considering the number of sections they have in common, but the duration they could, in theory, spend in common classrooms during a week. While a common one-credit-hour course section between students $i$ and $j$ would contribute the same as a common three-credit-hour course section in matrix $A$, the element $A_{weighted}[i,j]$ represents the total amount of time students $i$ and  $j$ would spend in common classrooms during a week, assuming there are no class absences.

\section{Formal Definition of Terms and Metrics}
\label{Metrics}

Consider $A_{binary}$, the matrix of student interactions in courses. As stated in Section~\ref{Adjacency_matrix}, if $A_{binary}[i,j]=1$, then students $i$ and $j$ have at least one course section in common. We also set $A_{binary}[i,i]=1$ because doing so helps in some calculations described later. We use this matrix to represent an undirected graph $G=(V,~E)$ with the following properties:
\begin{itemize}
\item $V$ is the set of vertices of the graph and contains one vertex for each student. The total number of vertices, $\mid V \mid$, is equal to $n$. Vertex $v_i: 1\leq i \leq n$ represents the $i^{th}$ student.

\item $E$ is the set of edges of the graph. There is an edge $(v_i,v_j) \in E$ if an only if $A_{binary}[i,j]=1$.
\end{itemize}


Let $s(i, j)$ denote the length of the \textit{shortest path} between vertices $v_i$ and $v_j$ in graph $G$. Note that there are $\binom{n}{2} = \frac{n(n-1)}{2}$ pairs of vertices in a graph of $n$ vertices. Therefore, the \textit{average path length}, $\ell_G$, of a connected, undirected graph $G$ of $n$ vertices (which refers to the mean of the shortest distances between all pairs of nodes) can be calculated as follows:
\begin{equation}
\ell_G = \frac{2}{n(n-1)}  \sum_{1\leq i<j\leq n} s(i,j)
\end{equation}

The \textit{diameter} of a connected graph $G$ is:
\begin{equation}
diameter_G = \max_{v_i,~v_j \in V} s(i,j)
\end{equation}
Low values of $diameter_G$ and $\ell_G$ for the student enrollment graph would be indicative of a small world network.

A vertex $v_i$ is said to be part of a \textit{triangle} if it has two neighbors $v_j$ and $v_k$ that are also neighbors of each other. In other words, the edges $(v_i,v_j)$, $(v_j,v_k)$ and $(v_k,v_i)$ form a triangle. If vertex $v_i$ has degree $d$, then $v_i$ can be part of up to $\binom{d}{2} = \frac{d(d-1)}{2}$ triangles. Let $T(v_i)$ represent the actual number of triangles that $v_i$ belongs to. The \textit{unweighted clustering coefficient}, $c(v_i)$, which indicates the cliquishness between vertex $v_i$ and its neighbors ($i.e.$, the fraction of all possible triangles involving $v_i$ that are actually present in the graph), can be calculated as follows:
\begin{equation}
c(v_i) = \frac{2T(v_i)}{d(d - 1)}
\end{equation}

The mean value of $c(v_i)$ over all vertices $v_i \in V$ is the \textit{average local clustering} coefficient, $C$, for the network:
\begin{equation}
C_G = \frac{1}{n} \sum_{v_i \in V} c(v_i)
\end{equation}

A set of three vertices that are reachable from each other is called a triad. A triad could either be a triangle or a pair of edges between three vertices.  The \textit{global clustering coefficient}, also referred to as transitivity, of a graph $G$ is expressed as:
\begin{equation}
T_G = \frac{3\times \textnormal{number of triangles}}{\textnormal{number of triads}} 
\end{equation}

The second power of $A_{binary}$ is represented as $A_{binary}^2 = A_{binary} \times A_{binary}$. If $A_{binary}^2[i,j]$ is non-zero, the shortest path between vertices $v_i$ and $v_j$ is at most two. Generalizing this notation, $A_{binary}^k = A_{binary}^{k-1} \times A_{binary}$, and if $A_{binary}^k[i,j]$ is non-zero, the shortest path between $v_i$ and $v_j$ is no more than $k$.\footnote{Note that setting $A_{binary}[i,i]$ to $1$ for all $i$ ensures that once $A^k_{binary}[i,j]$ becomes non-zero, it stays so for higher powers of $A_{binary}$.} The \textit{density} of an $n\times n$ adjacency matrix $A_{binary}$ raised to the power $k$ is the fraction of non-zero elements in $A_{binary}^k$:
\begin{equation}
\rho_{A_{binary}^k} = \frac{\textnormal{number of non-zero elements in } A_{binary}^k}{n^2}
\end{equation}
where $0\leq\rho\leq1$. This property measures the reachability of students within path length $k$ and may not reach $1$ for finite values of $k$ if the graph represented by $A$ is disconnected.

The network density $r$ is the ratio between the number of observed edges in an undirected graph $G$ and the total number of possible edges. For a graph with $m$ edges and $n$ vertices, it is expressed as:
\begin{equation}
r_G = \frac{2m}{n(n-1)}
\end{equation}
As $r$ approaches 1, $G$ gets closer to a complete graph. 

The \textit{betweenness centrality} of a node $v$ in a weighted, connected graph $G$ is
\begin{equation}
b_v = \sum_{s,t \in V} \frac{\sigma_{st}(v)}{\sigma_{st}}
\end{equation}
where $\sigma_{st}$ is the number of shortest paths between $s$ and $t$ and $\sigma_{st}(v)$ is the number of those paths that also pass through $v$. By this definition, if $v=s$ or $v=t$, then $\sigma_{st}(v)$ = 0, and if $s=t$, then $\sigma_{st}=0$. 

The betweenness centrality described above is a function of the number of vertices in the graph. So, to compare the betweenness centrality of vertices in graphs of different sizes, it is necessary to normalize the value as a function of graph size. For a connected graph of $n$ vertices, the maximum value of betweenness centrality of a vertex is $\binom{n-1}{2} = \frac{(n-1)(n-2)}{2}$, which arises when that vertex is the hub of a star network with $n-1$ neighbors. Hence, we report the \textit{normalized betweenness centrality} of a vertex as follows:
\begin{equation}
b_{v,norm} = \frac{2\times b_v}{(n-1)(n-2)}
\end{equation}

A student who is represented by a vertex with high betweenness centrality connects a large number of pairs of students. Such a student, if infected with a contagious disease, could be a proverbial \textit{superspreader}. Hence, in our analysis, we not only determine the betweenness centrality of vertices, but also try to identify the kinds of vertices that exhibit high betweenness centrality. Course delivery-based strategies to minimize rapid spread of infectious diseases should try to move the corresponding students to online instruction to minimize their physical interaction with other students.

Thus far, we have considered an unweighted graph of students where all edges have unit weight. Such a graph does not differentiate between a pair of students who are enrolled in a single common course that meets for an hour each week and another pair of students who are enrolled in three common courses, each of which meet for three hours per week. Hence, we define a new metric for weekly contact duration between students. Let $enrollment(x)$ be the set of students enrolled in course $x$, and let $contact(x)$ be the number of weekly class contact hours for course $x$. Then, the amount of time two students spend together in classrooms each week (assuming no absences) can be represented as:
\begin{equation}
contact\_duration(i, j) = \sum\{contact(x): i, j \in enrollment(x)\}
\end{equation}
Using this metric, we consider a weighted graph $G_{weighted}$ with the corresponding adjacency matrix $A_{weighted}$ such that $A_{weighted}[i,j] = \frac{1}{contact\_duration(i,j)}$. The more time students $i$ and $j$ spend together in class, the lower the weight of the edge between them. We determine minimal-weight paths between pairs of vertices and the betweenness centrality of vertices in this weighted graph. Our assumption is that the betweenness centrality metric for this weighted graph will be a better indicator of the potential for an infected student to spread a contagion through classroom interaction than the corresponding metric for an unweighted graph.



\section{Analysis of Data for Fall 2019}
\label{Analysis_Fall19}

The results of this study pertaining to the entire university are summarized in Table 1. A total of $28,849$ students registered in Fall 2019 courses, but the largest connected component of student-to-student network had only $27,080$ students. In this largest connected component of all students at the university, the average path length, $\ell_G$ (also referred to as the average geodesic distance), is equal to $2.97$. This implies that, on an average, two students are within a distance of three hops of each other. This short distance between students is more pronounced if we limit our analysis to undergraduate students for whom the average geodesic distance is a low $2.44$. Such a short distance in a large graph is indicative of a small-world network. 

\subsection{Graduate versus undergraduate level courses} 
Graduate students constitute $27\%$ of all students. Like the university-wide network, a majority ($79.1\%$) of the nodes in the graduate network belong to the main component, where defining small world characteristics emerged. The local clustering coefficient was much higher between graduate students ($0.63$) than the university wide network ($0.46$). In particular, we notice that unlike the interconnectivity between nodes of several majors in the university-wide network, the arrangement of nodes in the graduate student subnetwork involves students having more connections with their intra-school cohorts because they register in major-specific coursework. Figure 1c also reveals the distinct clustering of students in ECS, EPPS, AH, and SOM. Furthermore, NSM nodes cluster near the center, and many of these nodes act as bridges between major clusters, such as SOM and ECS.

The average geodesic distance in the graduate network is approximately 4.1. So, on average, connecting any two students in the graduate student network takes about 2 more steps than doing so requires in the undergraduate network. Although students may be more interconnected with other students of the same major, the formation of separate clusters may lead to an increased number of steps to connect students that may be in different majors, causing the average shortest path length to increase. 

Nodes in the graduate network have a significantly lower average degree ($94.5$) than those in the undergraduate network ($326.6$), indicating that, on average, graduate students have fewer direct connections to others. We also find that the average edge weight in the graduate network  is greater ($11.03$ credit hours) than that in the undergraduate network ($9.50$ credit hours). Taken together, these two findings explain that, although graduate students may be connected to fewer people on average, they spend more time in classes with each other compared to undergraduates.

The diameter of the largest connected component among graduate students, which corresponds to the distance between two farthest nodes, is only $9$. When limited to undergraduate students only, this number drops to $5$. So, more than $99\%$ of undergraduate students can reach each other within five steps.

As can be seen in the table, the largest connected component of students exhibits a high level of local ($0.46$) and global ($0.38$) clustering. Yet another indicator of the small world network is that such a high level of clustering is achieved with a relatively low level of network density, close to $0.01$.


\begin{table*}[!ht]
	
	\centering \small 
	\begin{tabular}{| l | r | r | r |}
		\hline
		Metric                              & University & Graduates & Undergraduates \\ \hline
		Nodes, \textit{full graph}          & 28,849      & 7,899      & 20,950          \\
		Edges, \textit{full graph}          & 3,714,254    & 298,316    & 3,394,312        \\
		Nodes, $n$                          & 27,080      & 6,252      & 20,784          \\
		Edges, $m$                          & 3,711,518    & 295,445    & 3,394,310        \\
		Average degree                      & 274.12     & 94.512    & 326.63         \\
		Percent nodes in largest comp.  & 93.861     & 79.149    & 99.208         \\
		Average edge weight                 & 9.6231     & 11.025    & 9.4966         \\
		Average geodesic distance, $\ell_G$ & 2.9694     & 4.1021    & 2.4425         \\
		Diameter of network                 & 9          & 10        & 5              \\
		Unweighted local $C_G$              & 0.46487    & 0.62702   & 0.41971        \\
		Unweighted global $T_G$             & 0.38386    & 0.45109   & 0.38247        \\
		Network density, $r_G$              & 0.010122   & 0.015120  & 0.015716       \\ \hline
	\end{tabular}
	\caption{Metrics for the largest connected component of student-to-student networks at the university-wide, graduate, and undergraduate levels.}
	\label{tab:metrics1}
\end{table*}

\subsection{Diving deeper: comparison between course levels}
\label{Diving_deeper}

Next, let us consider the relationships between students enrolled in all courses at specific levels (i.e., $1xxx$ through $7xxx$) as shown in Table~\ref{tab:metrics2}. As there are very few  $8xxx$ level courses, and they typically correspond to individual research credits taken by doctoral students, we do not report the corresponding numbers. Among undergraduate students, a low average geodesic distance between students is evident at all levels of courses ($1xxx$ through $4xxx$). These are markedly lower than the average geodesic distances for $5xxx$ and $6xxx$ level courses, which constitute the bulk of graduate level courses with classroom instruction. Also, the diameter (the longest path between any two nodes) for the $1xxx$ level courses is significantly lower ($5$) than it is at the $4xxx$ level courses ($8$), and the diameters at all four undergraduate course levels are lower than the diameter for $5xxx$ and $6xxx$ level courses ($11$). 

\begin{table*}[!ht]
    
    \centering \small
    \begin{tabular}{|l | r | r | r | r | r | r |}
    	\hline
    	                                    &             \multicolumn{6}{c|}{Course Catalog Prefix}             \\ \cline{2-7}
    	Metric                              & 1xxx     & 2xxx     & 3xxx      & 4xxx     & 5xxx, 6xxx & 7xxx     \\ \hline
    	Nodes, \textit{full graph}          & 9,149     & 12,454    & 13,285     & 8,454     & 7,346       & 738      \\
    	Edges, \textit{full graph}          & 879,833   & 1,346,377  & 866,593    & 435,992   & 305,919     & 10,167    \\
    	Nodes, $n$                          & 9,145     & 12,413    & 13,183     & 7,999     & 6,354       & 178      \\
    	Edges, $m$                          & 879,823   & 1,346,377  & 866,410    & 435,956   & 303,180     & 5,586     \\
    	Percent nodes in largest comp.      & 99.956   & 99.671   & 99.232    & 94.618   & 86.496     & 24.119   \\
    	Average edge weight                 & 7.5897   & 9.1007   & 10.175    & 10.436   & 10.983     & 6.3960   \\
    	Average geodesic distance, $\ell_G$ & 2.5027   & 2.50434  & 2.98580   & 3.3771   & 4.1291     & 2.1117   \\
    	Diameter of network                 & 5        & 5        & 6         & 8        & 11         & 6        \\
    	Unweighted local $C_G$              & 0.72779  & 0.71452  & 0.66008   & 0.758890 & 0.64202    & 0.88821  \\
    	Unweighted global $T_G$             & 0.54725  & 0.57823  & 0.52284   & 0.63947  & 0.45031    & 0.85617  \\
    	Network density, $r_G$              & 0.021043 & 0.017477 & 0.0099714 & 0.013629 & 0.015021   & 0.354600 \\ \hline
    \end{tabular}
\caption{Metrics for the largest connected component of student-to-student networks at the various course levels.}
    \label{tab:metrics2}
\end{table*}

Freshmen students are not limited to taking $1xxx$ level courses, sophomores are not limited to taking $2xxx$ level courses, and so on. For example, a senior may be enrolled in a $1xxx$ or $2xxx$ level course that satisfies a general core requirement. So, we analyze course registration data for students classified as freshmen, sophomores, juniors and seniors separately. Similar trends, including a high level of clustering, low diameter, and low average geodesic distance, are evident, as shown in Table~\ref{tab:metrics3}.
\begin{table*}[!ht]
    
    \centering \small
    \begin{tabular}{| l | r | r | r | r |}
    	\hline
    	Metric                              & Freshman & Sophomore & Junior   & Senior   \\ \hline
    	Nodes, \textit{full graph}          & 4,330     & 3,220      & 6,119     & 7,088     \\
    	Edges, \textit{full graph}          & 663,636   & 203,254    & 350,073   & 426,513   \\
    	Nodes, $n$                          & 4,277     & 3,211      & 6,092     & 6,985     \\
    	Edges, $m$                          & 663,608   & 203,249    & 350,071   & 426,509   \\
    	Average degree                      & 310.315  & 126.595   & 114.9281 & 122.1214 \\
    	Percent nodes in largest comp.      & 98.776   & 99.720    & 99.559   & 98.547   \\
    	Average edge weight                 & 8.2339   & 10.373    & 10.226   & 11.414   \\
    	Average geodesic distance, $\ell_G$ & 1.9951   & 2.3124    & 2.6136   & 2.7737   \\
    	Diameter of network                 & 6        & 5         & 5        & 6        \\
    	Unweighted local $C_G$              & 0.42119  & 0.44724   & 0.4261   & 0.4914   \\
    	Unweighted global $T_G$             & 0.39850  & 0.46357   & 0.4252   & 0.4656   \\
    	Network density, $r_G$              & 0.072571 & 0.039444  & 0.018868 & 0.017486 \\ \hline
    \end{tabular}
\caption{Metrics for the largest connected component of student-to-student networks for undergraduate freshmen, sophomores, juniors, and seniors.}
    \label{tab:metrics3}
    \end{table*}

\begin{figure*}[!htb]
    \centering
    \includegraphics[width=1.0\linewidth]{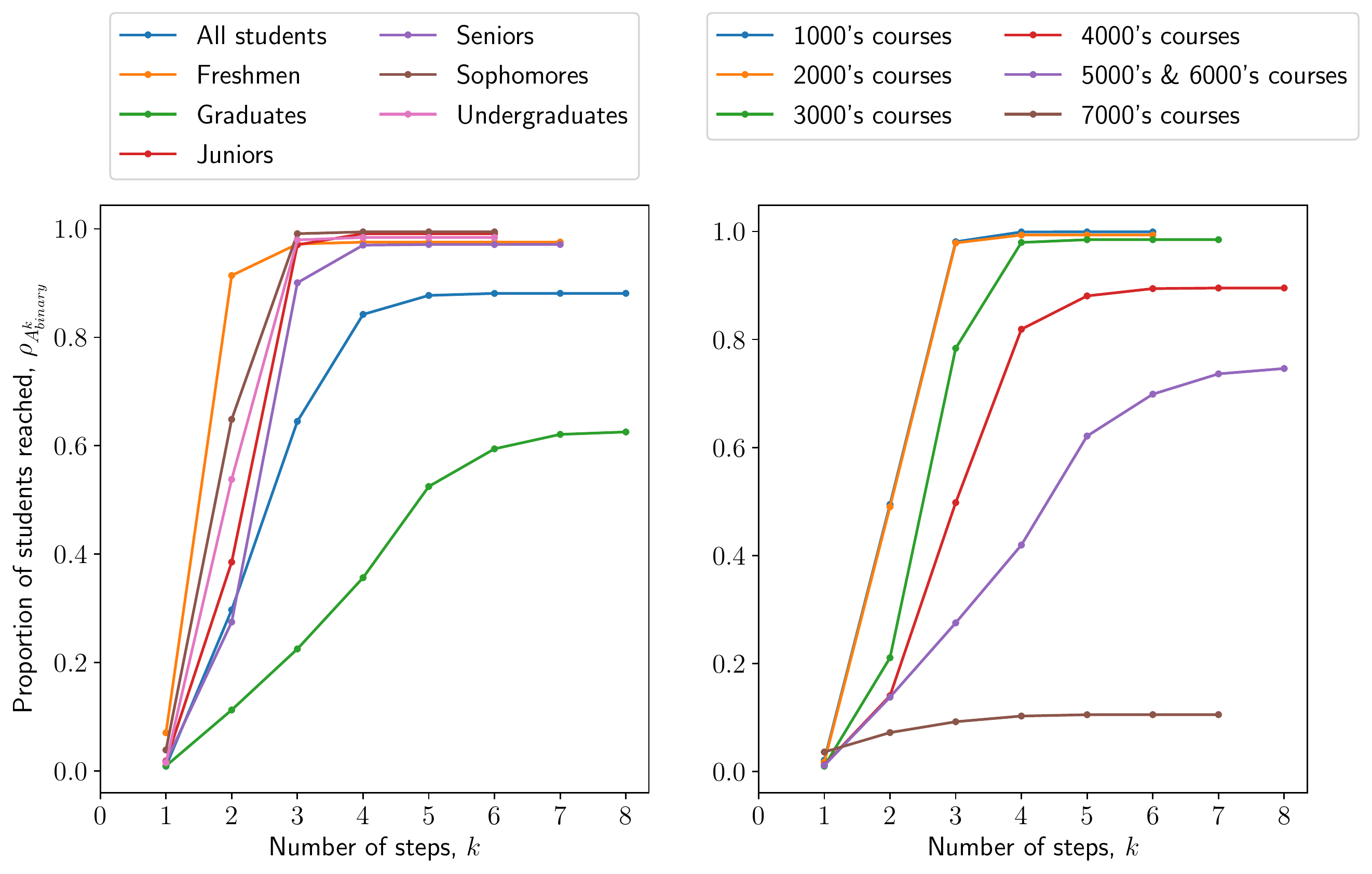}
    \caption{Cumulative density of matrices for consecutive powers of $A_{binary}$.}
    \label{fig:cumulative-dens}
\end{figure*}

Figure~\ref{fig:cumulative-dens} shows the cumulative distribution of student pairs that are reachable from each other within a given distance. Let us first consider the graph on the left. As seen in the plot for all students (blue curve), more than $80\%$ of student pairs are within a distance of $4$ of each other. Freshmen students are even more strongly interconnected (orange curve), with almost all students pairs within a distance of $4$ of each other (more than $90\%$ of freshmen students pairs are within a distance of $2$ of each other). Graduate students are not as closely related (green curve), with less than $40\%$ of them within a distance of $4$ of each other. Even when we consider a greater distance, interconnectivity remains somewhat low; only $60\%$ of graduate student pairs are within a distance of $6$ of each other. Similar trends are visible when we analyze the data for different course levels (i.e., $1xxx$ through $7xxx$; see graph on the right). The plots for the $1xxx$ and $2xxx$ courses are almost identical and appear as one (orange curve). Especially noteworthy, as an outlier, is the plot for $7xxx$-level courses, which indicates a high level of fragmentation for advanced graduate-level courses.

\begin{figure*}[!htb]
    \centering
    \includegraphics[width=1.0\linewidth]{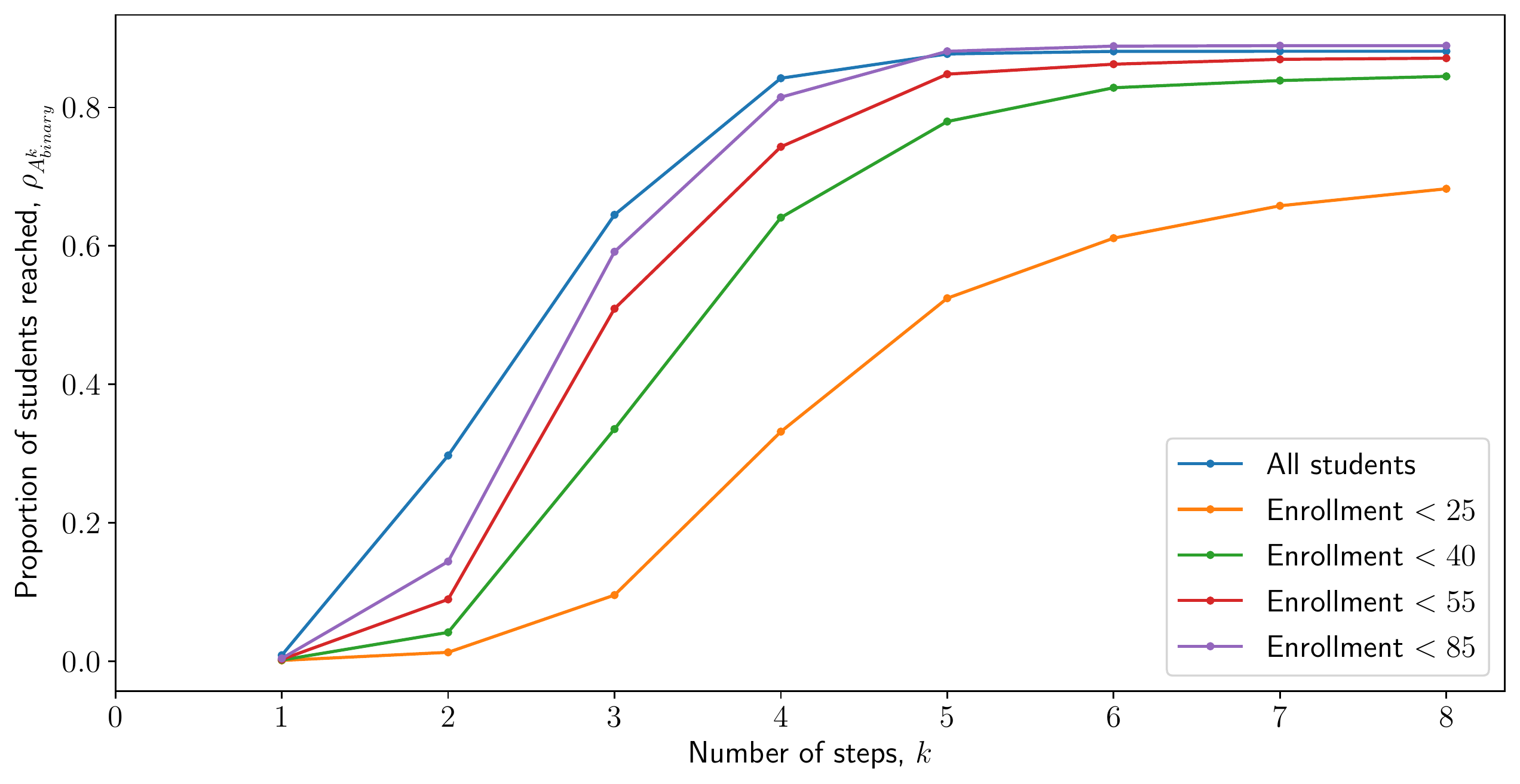}
    \caption{Cumulative density of matrices for consecutive powers of $A_{binary}$, keeping courses by size.}
    \label{fig:cumulative-dens-class-size}
\end{figure*}

Weeden and Cornwell \cite{Weeden_Cornwell_2020} investigate the impact of moving large enrollment courses online on interconnection between students taking in-person classes. They studied the impact of removing courses with enrollment over 30, 40, 50, 75 and 100 from the network on path lengths between students. We performed a similar simulation. The results are shown in Figure~\ref{fig:cumulative-dens-class-size}. Instead of choosing arbitrary thresholds, we relied on the quintile thresholds for course enrollment. Sections with enrollment less than $25$ account for $20\%$ of student enrollment in classes (first quintile). Sections with enrollment less than $40$ account for $40\%$ of class enrollments (second quintile). Similarly, the third and fourth quintile thresholds are sections with sizes of $55$ and $85$, respectively. 

Removing all sections with enrollment of $40$ or more, over $65\%$ of student pairs are within a distance of $4$ and more than $80\%$ are within a distance of $8$. When all sections with enrollment of $25$ or more are removed from the network, almost $40\%$ of student pairs are still within a distance of $4$ of each other and over $60\%$ are within a distance of $8$.





\subsection{Academic silos or strong interconnections}
\label{Silos}

Let us now dive deeper and consider the connectivity between individual students. Whereas Weeden and Cornwell \cite{Weeden_Cornwell_2020} considered a dual-mode graph of courses and students, we chose to focus on a single mode graph among students. The student-to-student network is drawn using the force-directed Fruchterman-Reingold algorithm. The more edge connections (shared courses) that two nodes (students) have in common, the closer together they are drawn.

\begin{figure*}[!htbp]
    \centering
    \begin{subfigure}[b]{1\linewidth}
        \includegraphics[width=\linewidth]{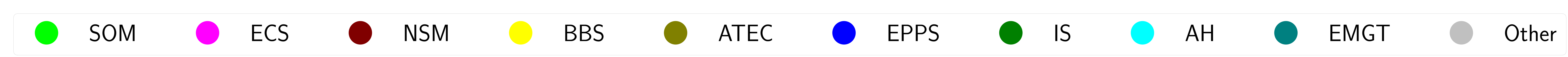} 
    \end{subfigure}
    \begin{subfigure}[b]{0.32\linewidth}
        \includegraphics[width=\linewidth]{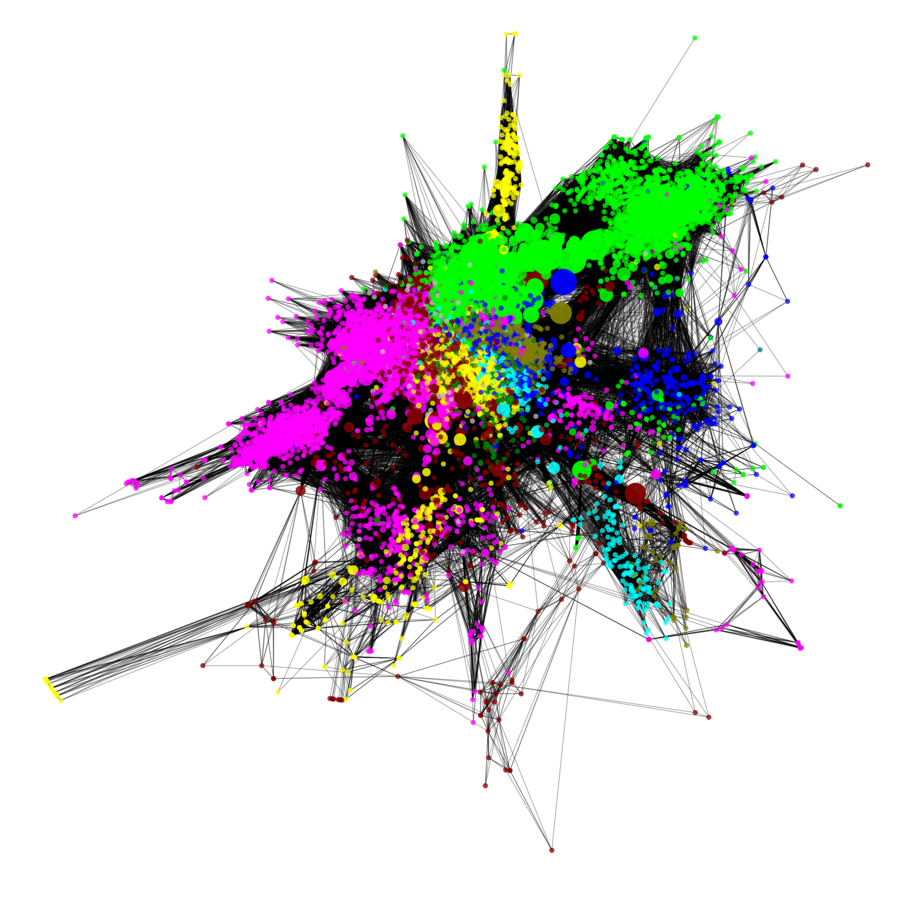} 
        \caption{University-wide network}
        \label{fig:graph-full-net}
    \end{subfigure} 
    \begin{subfigure}[b]{0.32\linewidth}
        \includegraphics[width=\linewidth]{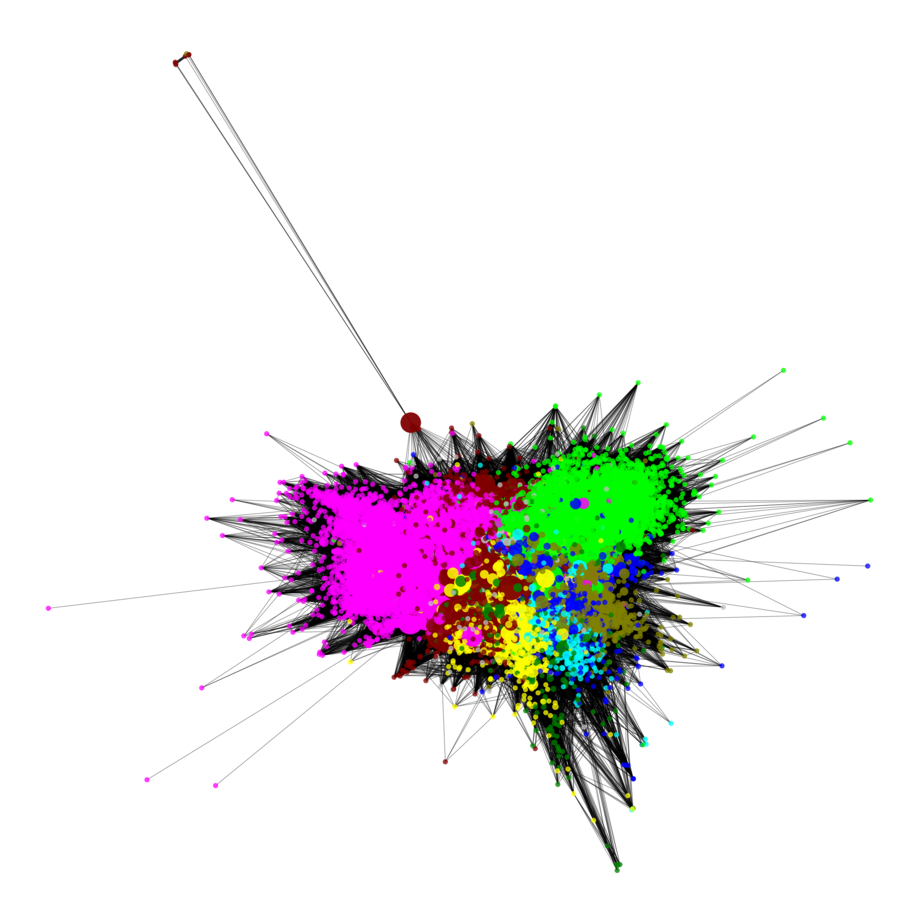}
        \caption{Undergraduate network}
        \label{fig:graph-ugrd-net}
    \end{subfigure}
    \begin{subfigure}[b]{0.32\linewidth}
        \includegraphics[width=\linewidth]{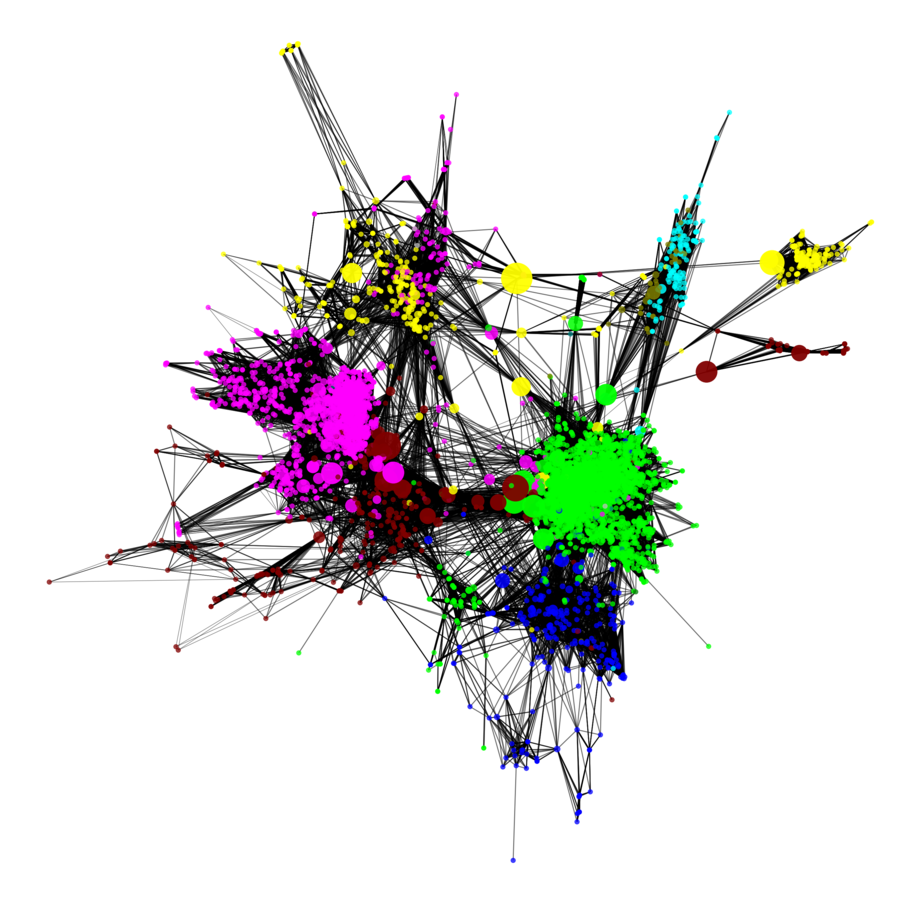}
        \caption{Graduate network}
        \label{fig:graph-grad-net}
    \end{subfigure}
    \caption{University wide, undergraduate, and graduate networks. Larger nodes have higher weighted betweenness centralities. }
    \label{fig:primary-networks}
\end{figure*}

Nodes in Figures~\ref{fig:primary-networks}, ~\ref{fig:lower-division-networks}, and ~\ref{fig:upper-division-networks} were colored according to the school or concentration that their corresponding students were primarily registered with: Management (SOM, green), Engineering \& Computer Science (ECS, pink), Natural Science \& Mathematics (NSM, maroon), Behavior \& Brain Sciences (BBS, yellow), Arts/Technology/Emerging Communications (ATEC, dark yellow), Economic/Political and Policy Sciences (EPPS, blue), Interdisciplinary Studies (IS, dark green), Arts \& Humanities (AH, cyan), Executive Management (EMGT, teal), and unspecified (grey). Though Executive Management courses are offered by the School of Management, students in these courses tend to form their own tight knit community for the most part and interact very little with other Management students.  A very small number of students fall into the unspecified category.

The size of nodes scales linearly relative to the value of the highest betweenness centrality and students that share at least one course are connected by a black edge. Edge thickness represents the number of course hours shared by the student-pair. Although the positions of individual nodes might not provide meaningful information, the relative positions of distinct regions in the network convey the interdisciplinary relationship of node clusters.

In Figure~\ref{fig:graph-full-net} we see that all students are closely connected, even though students belonging to the two largest schools by enrollment (ECS and SOM) tend to be clustered together. Looking at Figures~\ref{fig:graph-ugrd-net} and ~\ref{fig:graph-grad-net}, it can be inferred that graduate students have a higher level of clustering than undergraduate students, likely caused by increased registration in major-specific coursework. We also find that interdisciplinary relationships are visible through the relative distances of the clusters, especially between the AH, NSM, and ECS clusters, and that the NSM and ECS nodes are influential bridges between several clusters. 

\subsubsection{Lower division courses}

\begin{figure*}[!htbp]
	\centering
    \begin{subfigure}[b]{1\linewidth}
        \includegraphics[width=\linewidth]{legend.pdf} 
    \end{subfigure}
	\begin{subfigure}[b]{0.47\linewidth}
		\includegraphics[width=\linewidth]{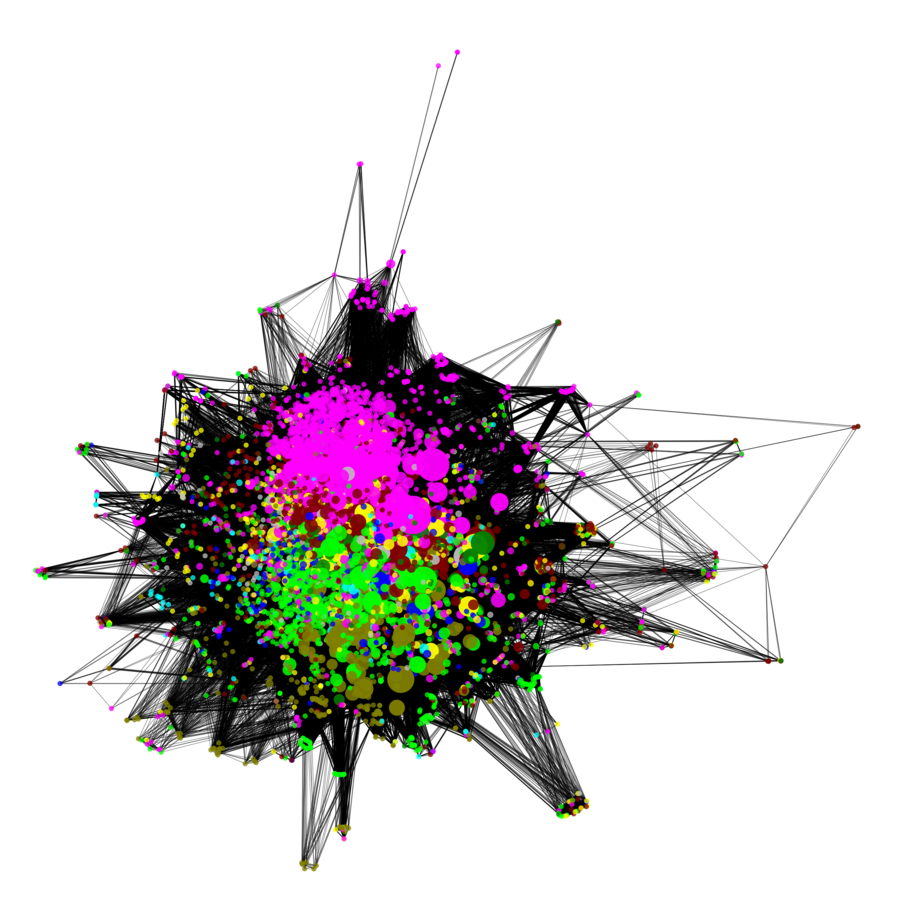}
		\caption{1000's level courses}
		\label{fig:graph-cat1}
	\end{subfigure}
	\begin{subfigure}[b]{0.47\linewidth}
		\includegraphics[width=\linewidth]{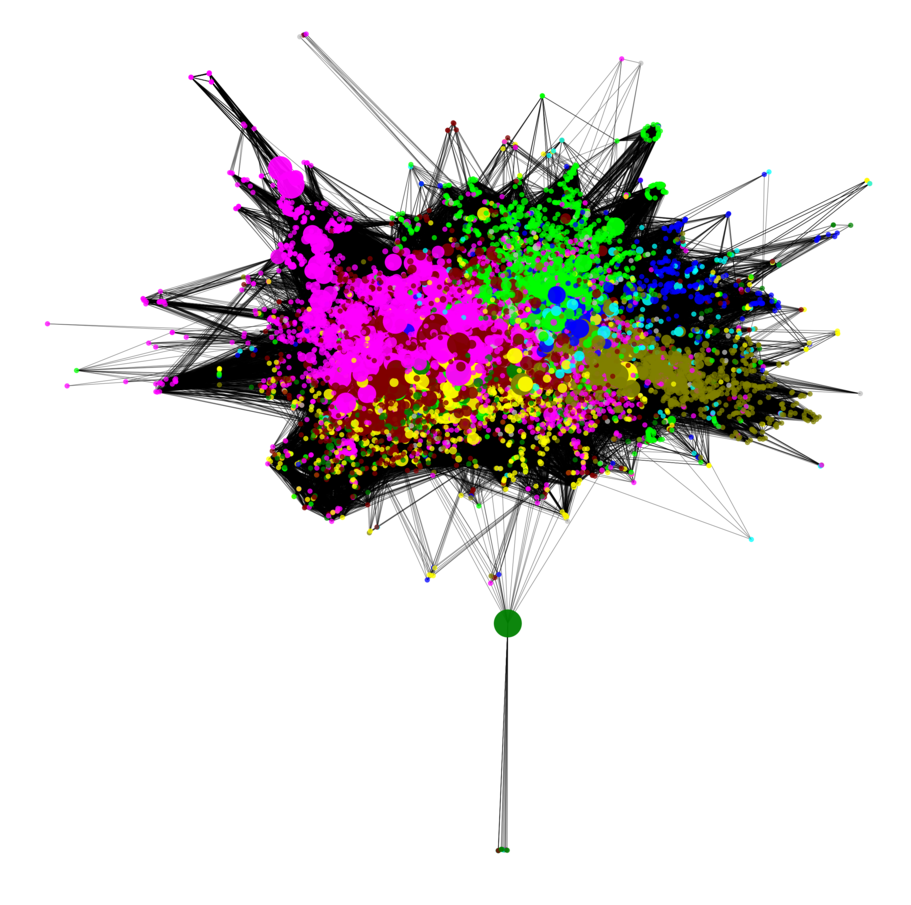}
		\caption{2000's level courses}
		\label{fig:graph-cat2}
	\end{subfigure}
	\caption{Lower division student-to-student networks partitioned by course level.  }
	\label{fig:lower-division-networks}
\end{figure*}

Figure~\ref{fig:lower-division-networks} shows the interconnections among students enrolled in lower-division undergraduate courses, $i.e.$, $1xxx$ and $2xxx$ level courses. As in the previous figure, larger sized nodes have higher weighted betweenness centralities. The graphs are clumped edges often connect students from different majors since many individuals share lower-level introductory classes. SOM and ECS dominate the graphs because they are the two most popular schools of study.

\subsubsection{Upper division courses}

\begin{figure*}[!htbp]
	\centering
    \begin{subfigure}[b]{1\linewidth}
        \includegraphics[width=\linewidth]{legend.pdf} 
    \end{subfigure}
	\begin{subfigure}[b]{0.47\linewidth}
		\includegraphics[width=\linewidth]{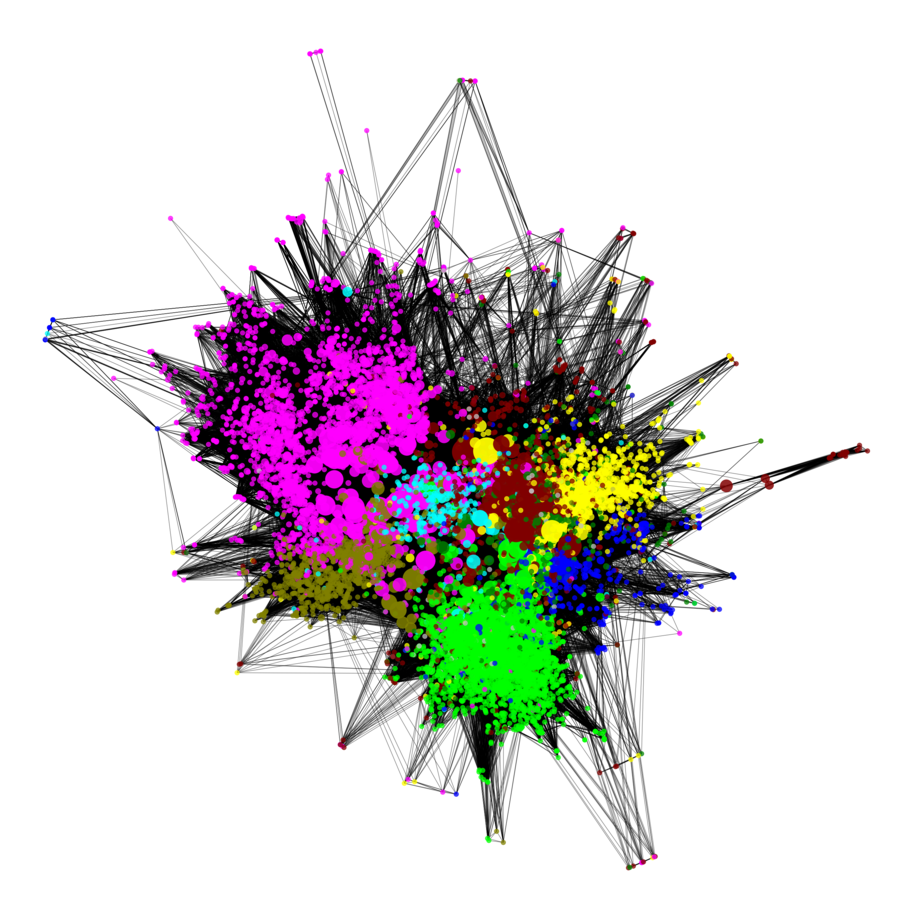}
		\caption{3000's level courses}
		\label{fig:graph-cat3}
	\end{subfigure}
	\begin{subfigure}[b]{0.47\linewidth}
		\includegraphics[width=\linewidth]{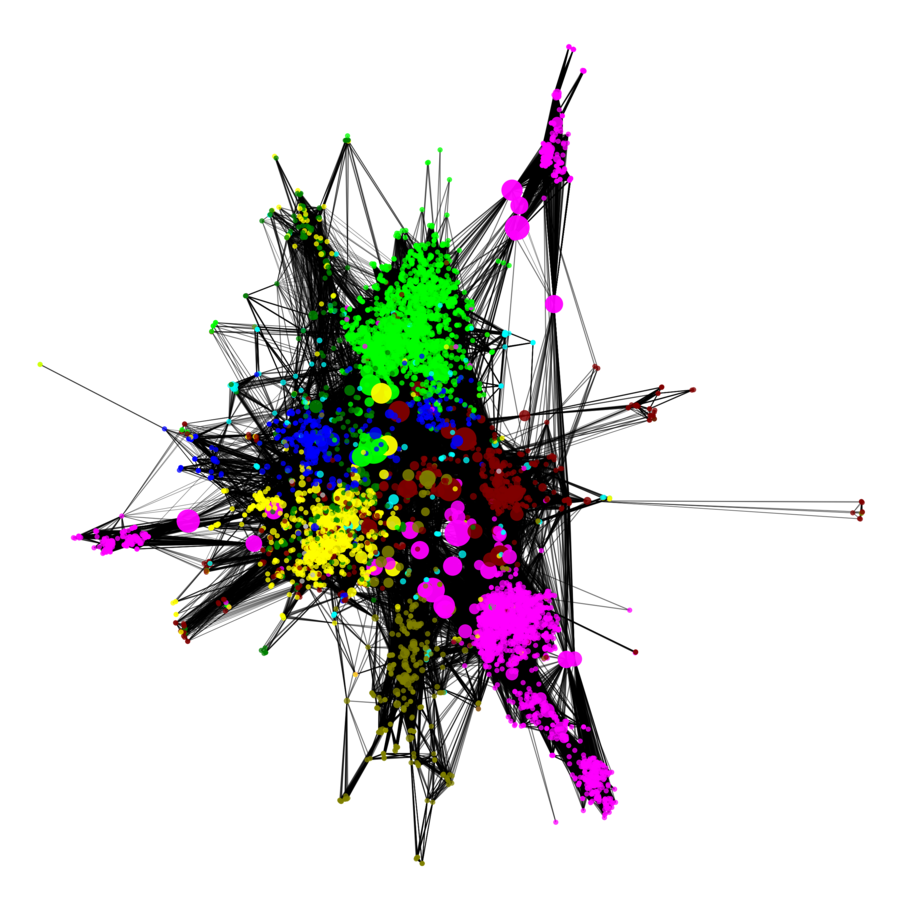}
		\caption{4000's level courses}
		\label{fig:graph-cat4}
	\end{subfigure}
	\begin{subfigure}[b]{0.47\linewidth}
		\includegraphics[width=\linewidth]{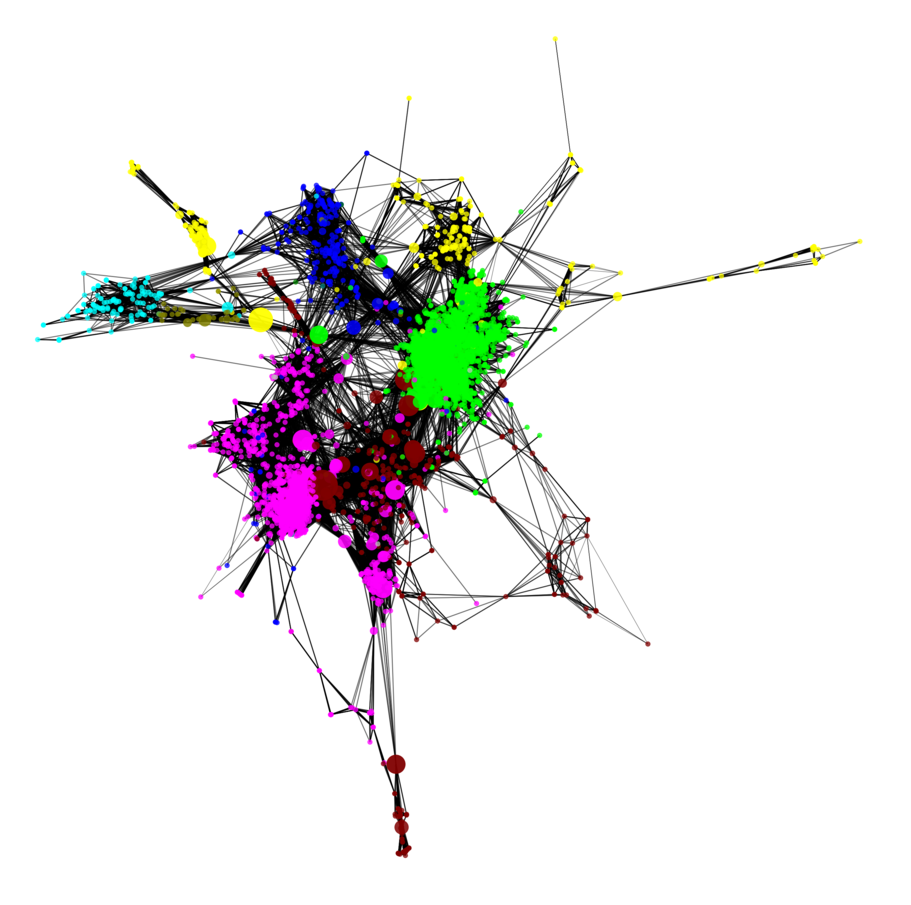}
		\caption{5000's and 6000's level courses}
		\label{fig:graph-cat56}
	\end{subfigure}
    \begin{subfigure}[b]{0.47\linewidth}
        \includegraphics[width=\linewidth]{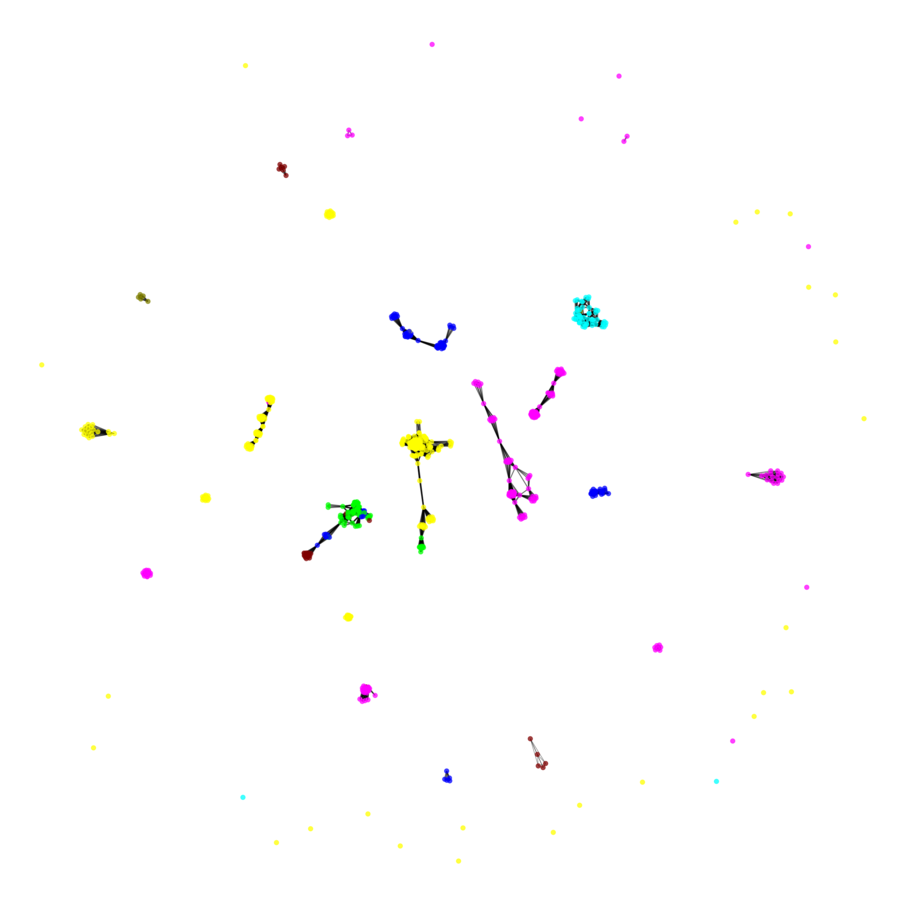}
        \caption{7000's level courses (full network is shown)}
        \label{fig:graph-cat7}
    \end{subfigure}
    \caption{Upper division student-to-student networks partitioned by course level. }
	\label{fig:upper-division-networks}
\end{figure*}

Figure~\ref{fig:upper-division-networks} shows the connections between students taking upper division undergraduate courses ($3xxx$ and $4xxx$), classroom-based graduate level courses ($5xxx$ and $6xxx$), and graduate level seminar and independent study courses ($7xxx$). Once again, larger nodes have higher weighted betweenness centralities. Clusters become increasingly visible in higher-level courses and they begin to rely more on bridge nodes to connect clusters. At the $7xxx$ level, clusters break down into components that mostly specialize in a specific field of study; only about 24\% of nodes are in the largest component. As the $7xxx$ level courses tend to be populated primarily by doctoral and master's thesis students, the significant level of intra-program clustering and disconnections between programs indicate that these courses could be the first ones to be conducted in-person due to the low risk of spreading an infectious disease, compared to undergraduate courses. 

\begin{figure*}[!htbp]
    \centering
    \begin{subfigure}[b]{1\linewidth}
        \includegraphics[width=\linewidth]{legend.pdf} 
    \end{subfigure}
    \begin{subfigure}[b]{0.47\linewidth}
        \includegraphics[width=\linewidth]{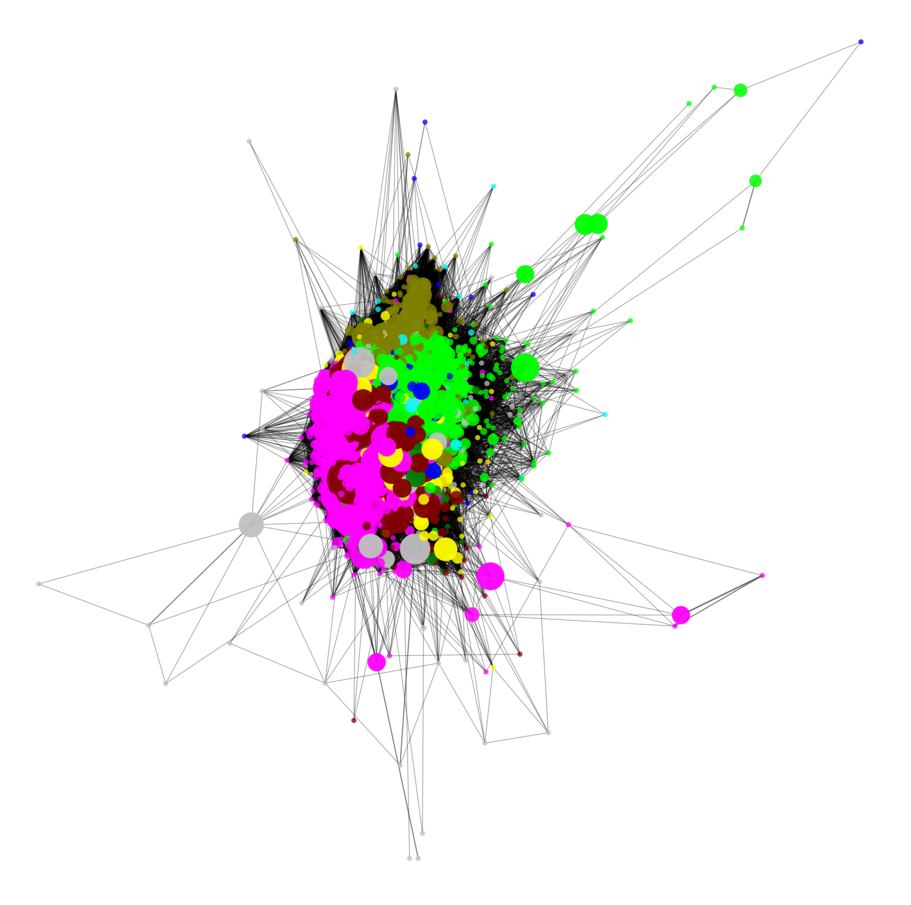}
        \caption{Freshman level network}
        \label{fig:graph-fresh}
    \end{subfigure}
    \begin{subfigure}[b]{0.47\linewidth}
        \includegraphics[width=\linewidth]{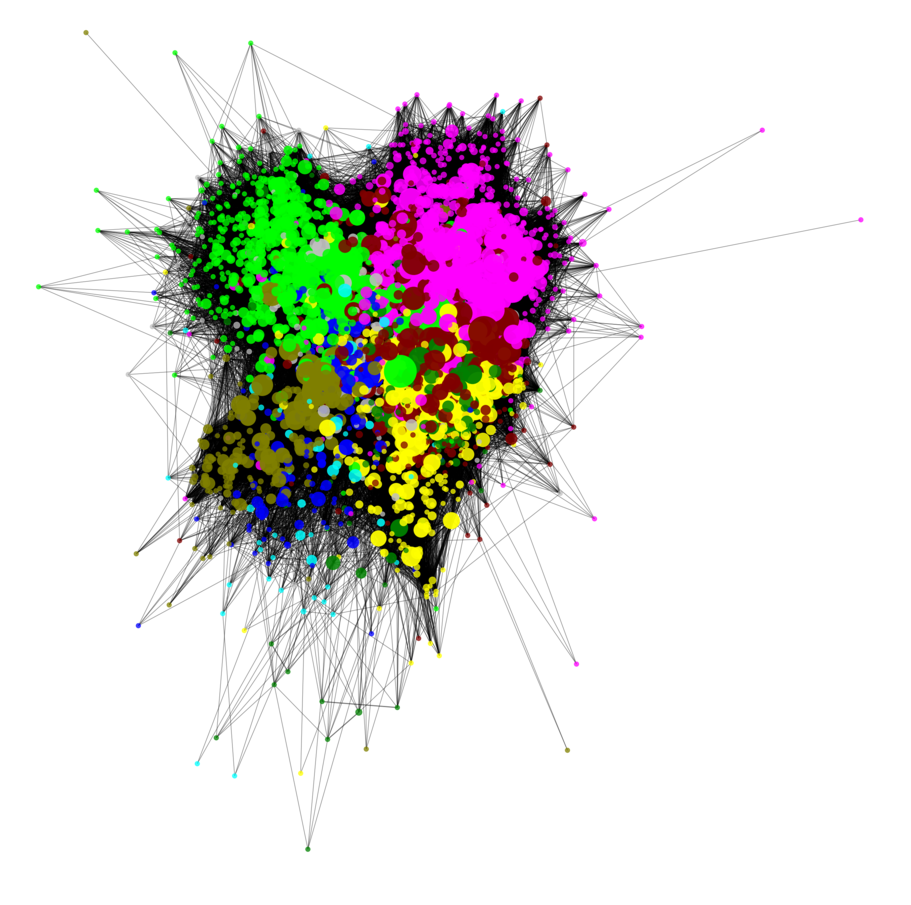}
        \caption{Sophomore level network}
        \label{fig:graph-soph}
    \end{subfigure}
    \begin{subfigure}[b]{0.47\linewidth}
        \includegraphics[width=\linewidth]{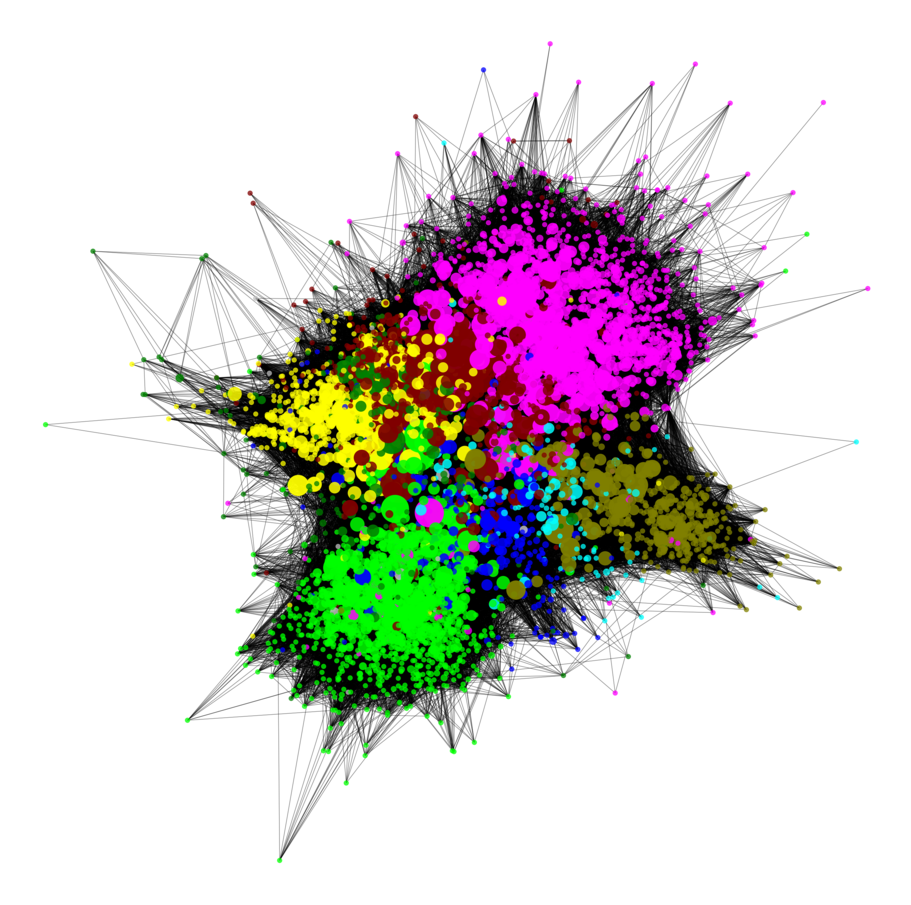}
        \caption{Junior level network}
        \label{fig:graph-jun}
    \end{subfigure}
    \begin{subfigure}[b]{0.47\linewidth}
        \includegraphics[width=\linewidth]{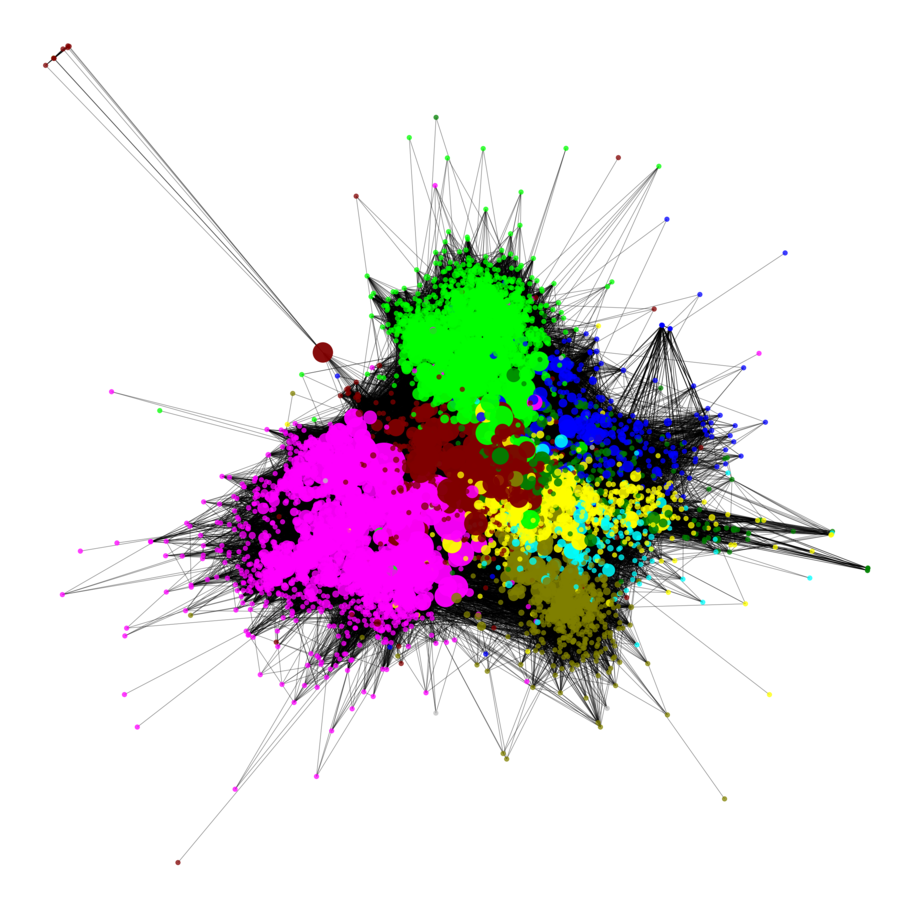}
        \caption{Senior level network}
        \label{fig:graph-sen}
    \end{subfigure}
    \caption{Student-to-student networks partitioned by student rank (freshmen, sophomores, juniors, and seniors). Larger nodes have higher weighted betweenness centralities. }
    \label{fig:student-rank-networks}
\end{figure*}



\subsection{Courses contributing to high betweenness centrality}
\label{Courses_pivotal}

As shown in Figure~\ref{fig:cumulative-dens-class-size}, even if all sections with enrollment of $25$ or more are removed from the network, a significant number of students are still able to reach each other through short paths. Hence, we investigated another possibility of disrupting the small world network between students: eliminating courses that contribute to high betweenness centrality. 

As betweenness centrality in a student-to-student network is a property of individual nodes, we investigated this idea as follows. For a given set of students, first we identified the \textit{one hundred} students with the highest betweenness centrality. We refer to them as the \textit{pivotal students}. Then, we identified all sections in which these pivotal students were enrolled. This simulation was conducted on the unweighted network of students as well as the weighted network described in Section~\ref{Metrics}. Our conjecture is that removing classes with high pivotal student enrollment from the student-to-student network ($i.e.$, moving these courses online) could disrupt the small world network, increase average path length between student pairs, and also increase the diameter of the network.

\begin{figure*}[!htbp]
    \centering
    \begin{subfigure}[b]{0.67\linewidth}
        \includegraphics[width=\linewidth]{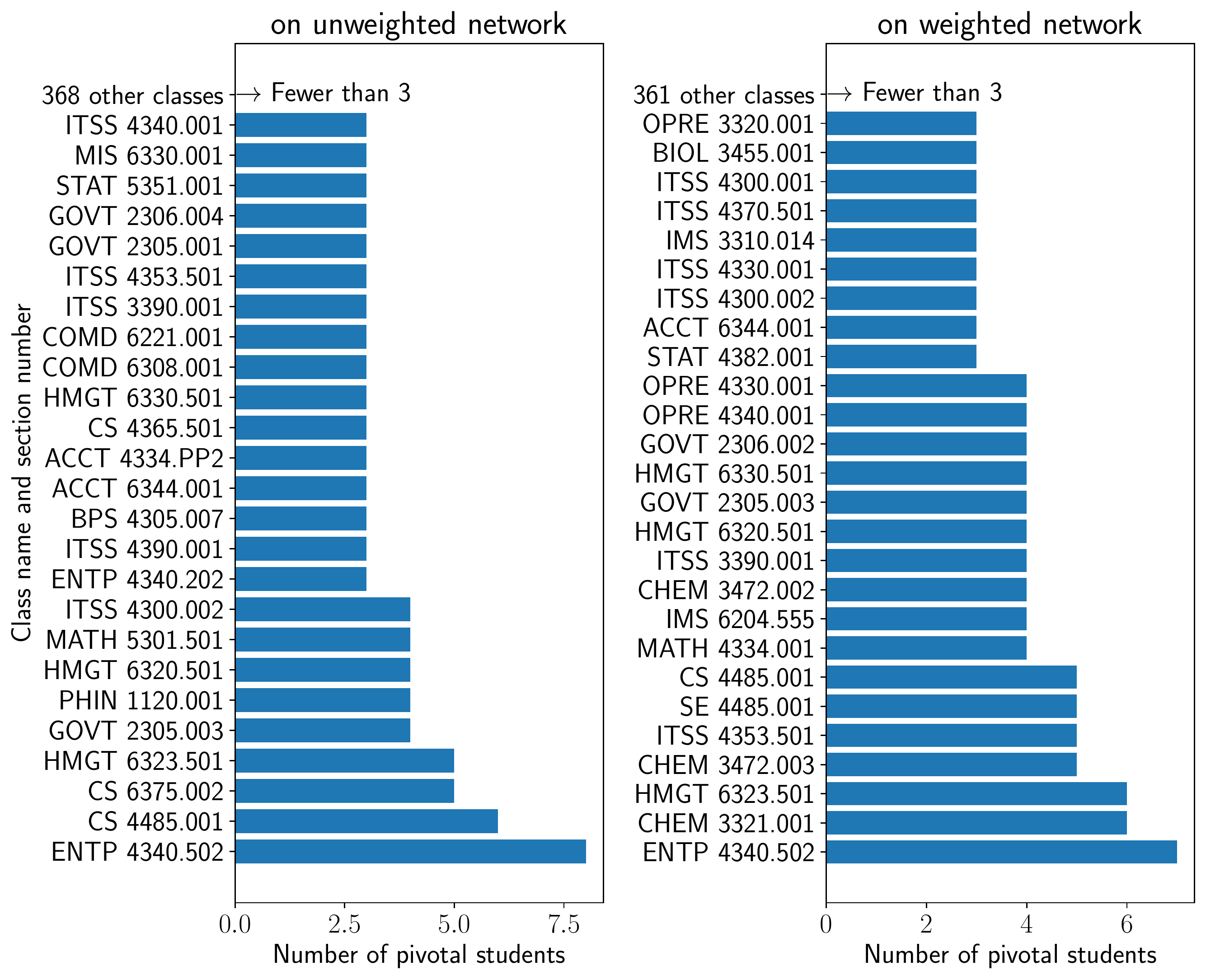}
        \caption{University-wide network}
        \label{fig:centrality-all}
    \end{subfigure}\\
    \begin{subfigure}[b]{0.49\linewidth}
        \includegraphics[width=\linewidth]{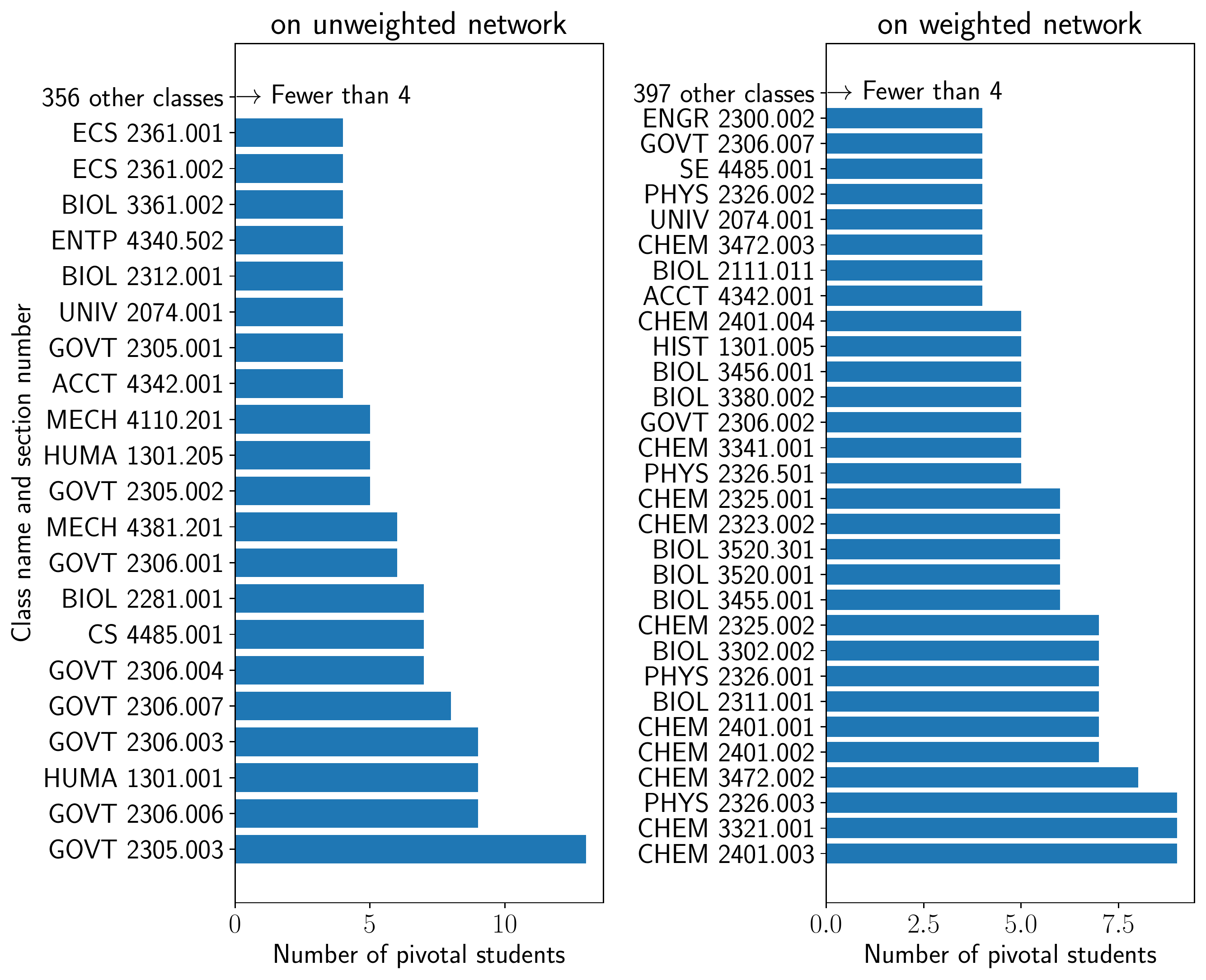}
        \caption{Undergraduate students only}
        \label{fig:centrality-undergrad}
    \end{subfigure}
    \begin{subfigure}[b]{0.49\linewidth}
        \includegraphics[width=\linewidth]{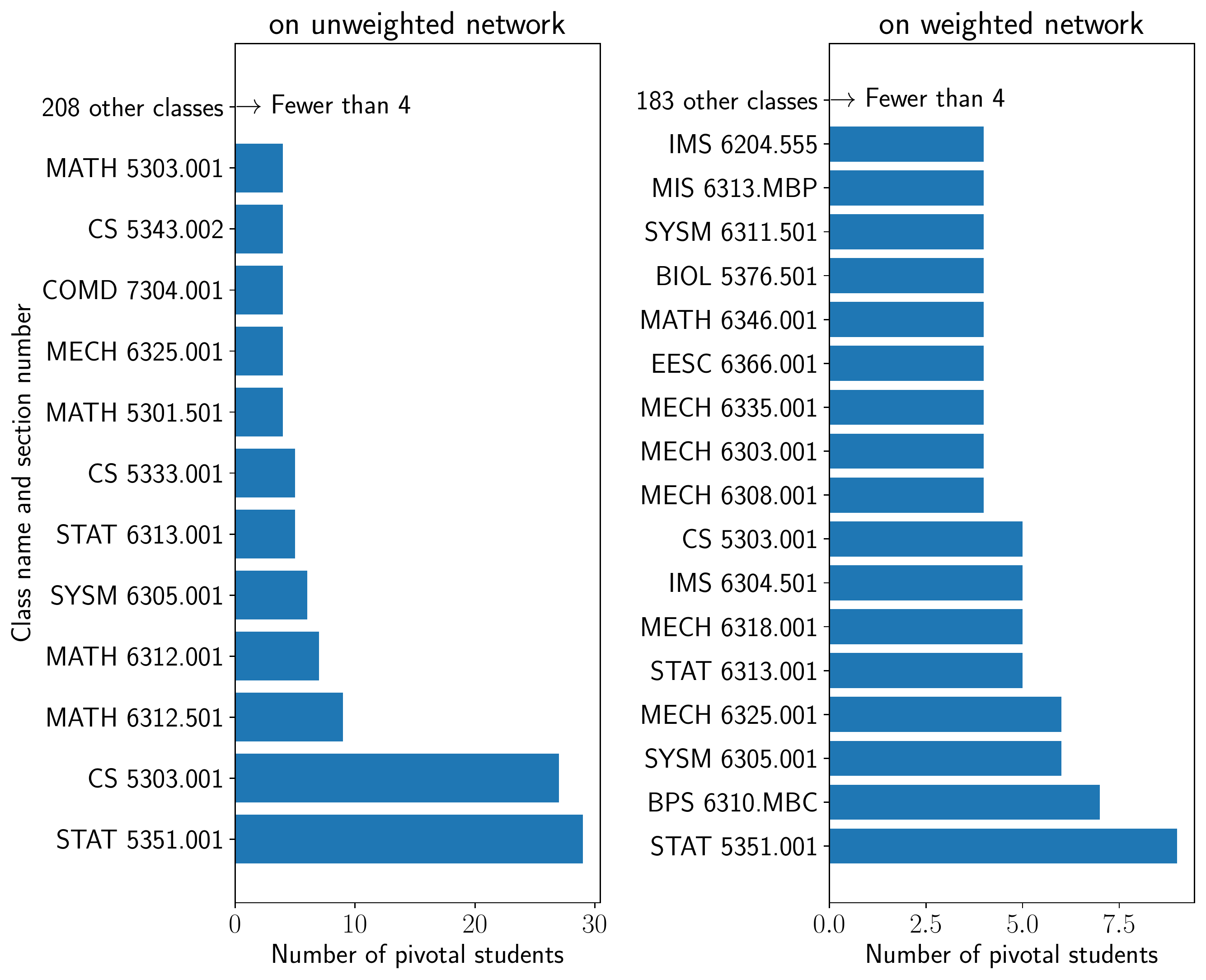}
        \caption{Graduate students only}
        \label{fig:centrality-grad}
    \end{subfigure}
    
    \caption{The classes most frequently taken by pivotal students (nodes with high betweenness centralities) in student-to-student networks.}
    \label{fig:btw}
\end{figure*}

Figure~\ref{fig:btw} lists sections in which pivotal students are enrolled. Figure~\ref{fig:centrality-all} shows such courses for the entire student population, while Figure~\ref{fig:centrality-undergrad} and ~\ref{fig:centrality-grad} are limited to undergraduate and graduate students, respectively. Courses shown in Figure~\ref{fig:btw} are not surprising. They either correspond to upper division undergradaute or graduate courses (e.g., ENTP4340, CS 4485, CS 6375) from ECS and SOM, the two schools with largest enrollment; or they correspond to lower division courses that satisfy the general core requirement (e.g., GOVT 2305, GOVT 2306). Eight out of the one hundred students with the highest betweenness centralities are enrolled in ENTP 4340.

An interesting situation emerges when we limit the analysis to undergraduate (Figure~\ref{fig:centrality-undergrad}) or graduate (Figure~\ref{fig:centrality-grad}) students. While eight students taking ENTP 4340 were among the one hundred nodes with highest betweenness centrality for the university-wide network, only four of them are among the top one hundred when the analysis was limited to undergraduate students. The other four most likely had high betweenness centralities because they were also taking some graduate level courses and were on the shortest paths between several graduate and undergraduate students. Those links disappeared when the ananlysis was limited to undergraduate students. Similarly, only three of the one hundred students with the highest betweenness centralities were in GOVT 2305. However, when the analysis was limited to undergraduate students, that number jumped to thirteen. 

\begin{figure*}[!htbp]
    \centering
    \begin{subfigure}[b]{0.49\linewidth}
        \includegraphics[width=\linewidth]{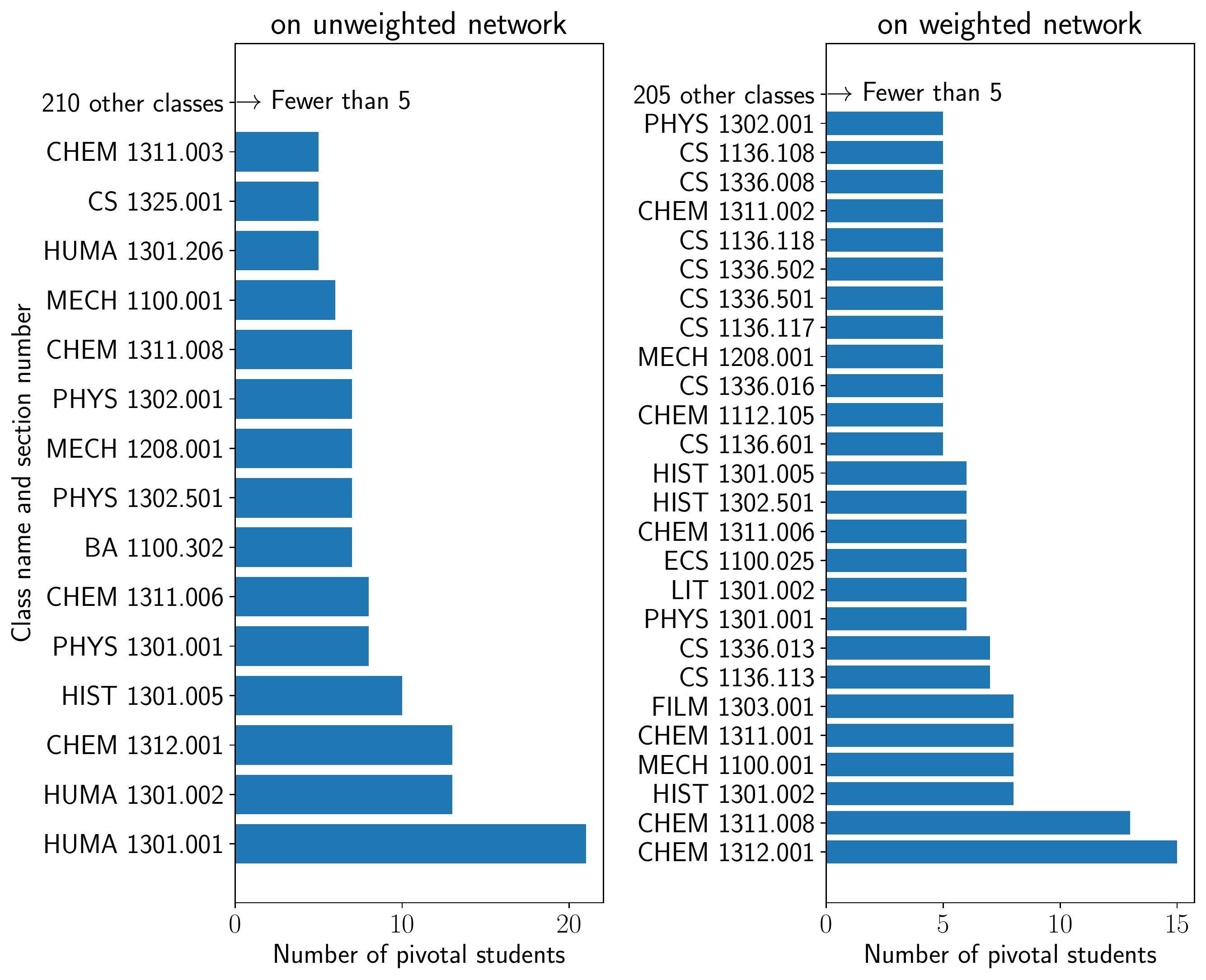}
        \caption{1000's level courses}
        \label{fig:centrality-cat1}
    \end{subfigure}
    \begin{subfigure}[b]{0.49\linewidth}
        \includegraphics[width=\linewidth]{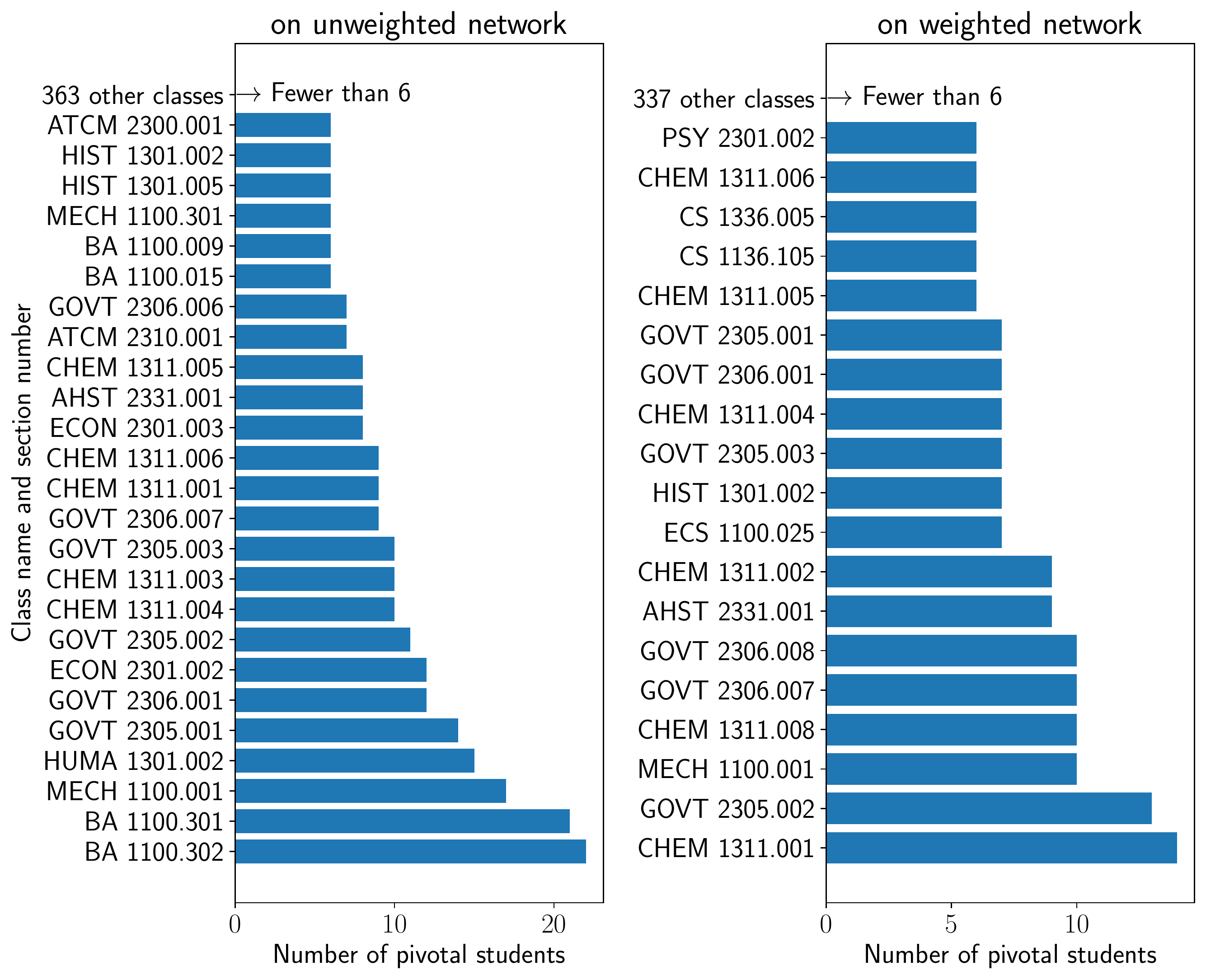}
        \caption{Freshmen level students only}
        \label{fig:centrality-fresh}
    \end{subfigure}
    \begin{subfigure}[b]{0.49\linewidth}
        \includegraphics[width=\linewidth]{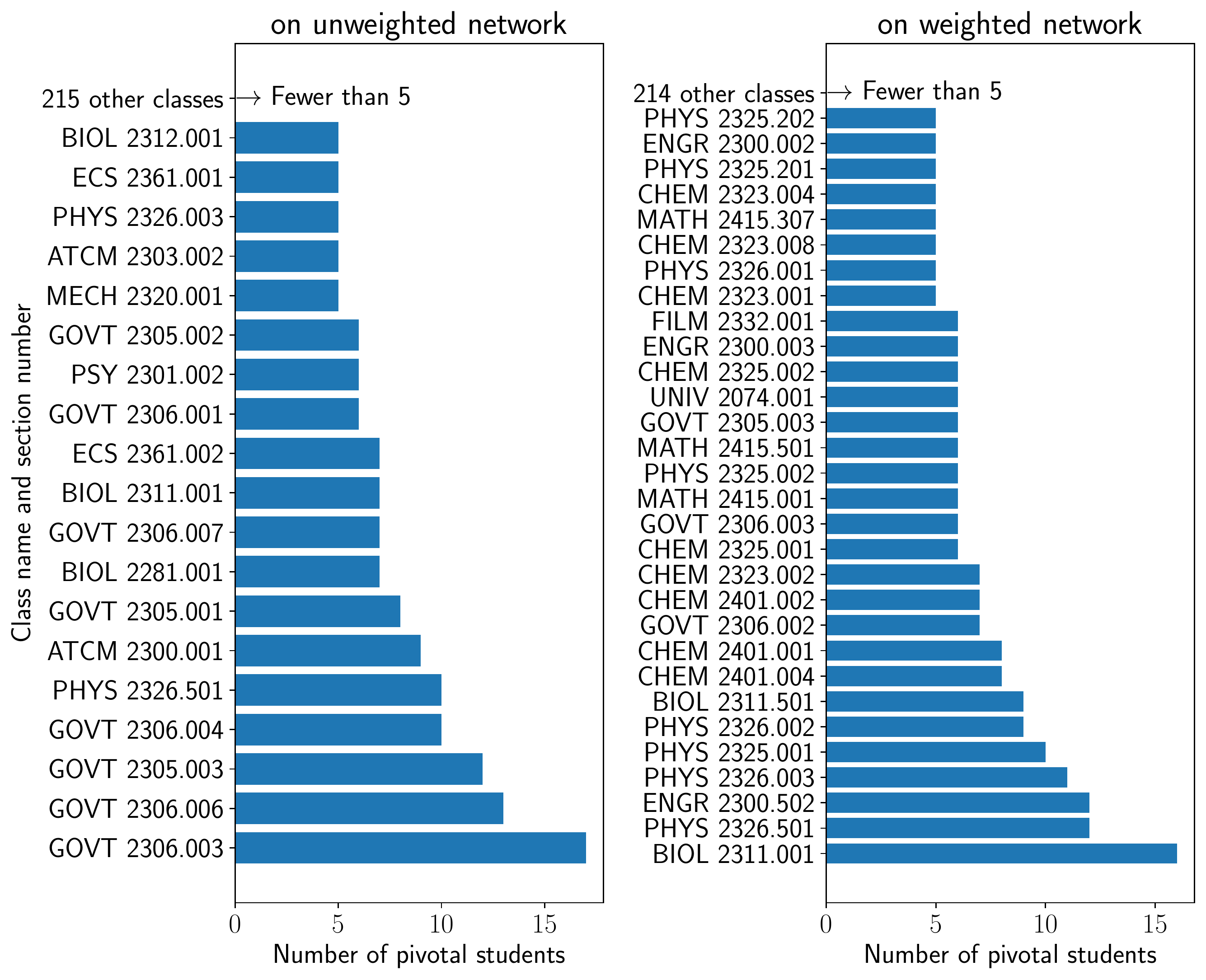}
        \caption{2000's level courses}
        \label{fig:centrality-cat2}
    \end{subfigure}
    \begin{subfigure}[b]{0.49\linewidth}
        \includegraphics[width=\linewidth]{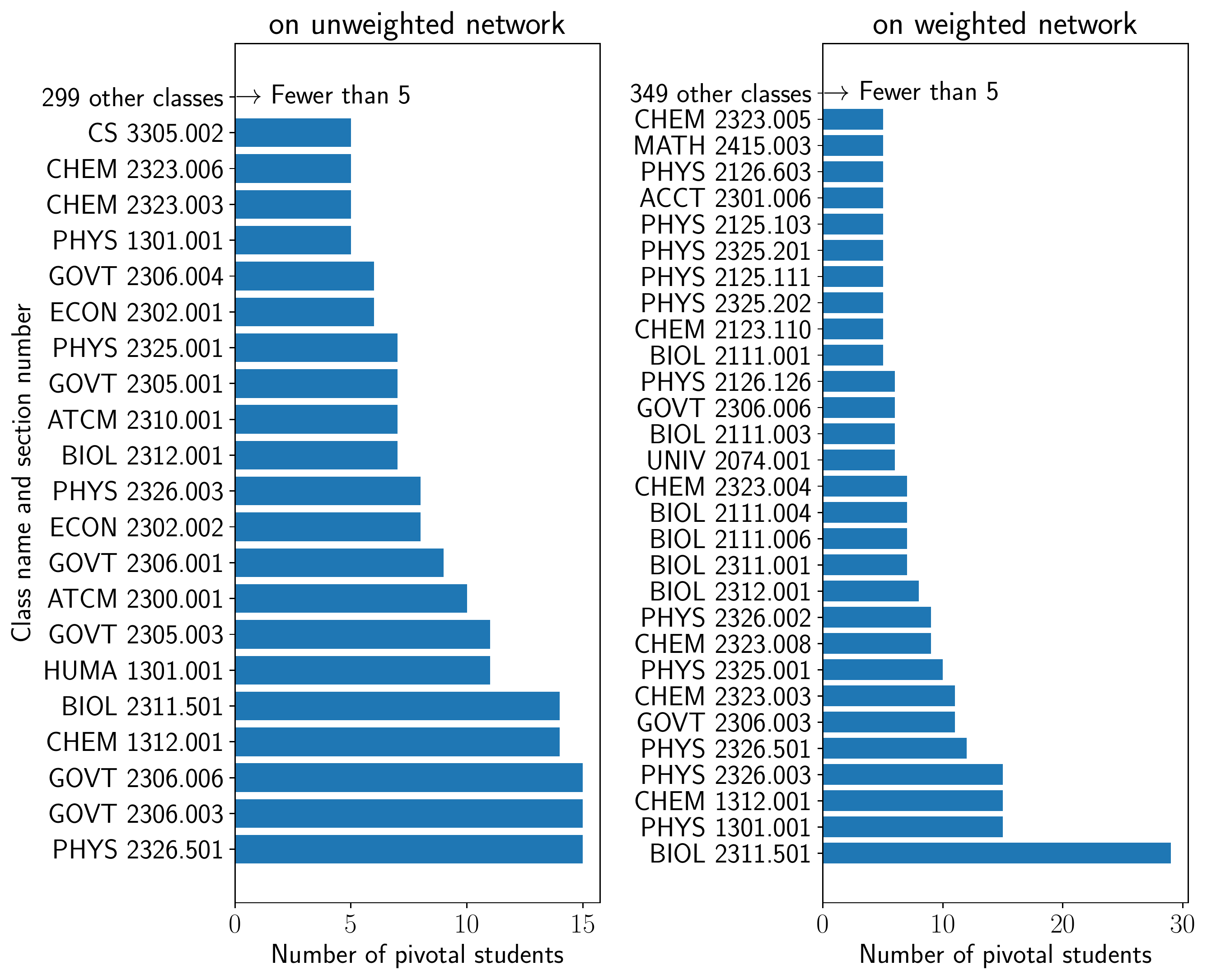}
        \caption{Sophomore level students only}
        \label{fig:centrality-soph}
    \end{subfigure}
    
    \caption{The classes most frequently taken by pivotal students (nodes with high betweenness centralities) in lower division student-to-student networks.}
    \label{fig:btw-lower-division}
\end{figure*}

Figure~\ref{fig:btw-lower-division} shows the courses taken by the one hundred pivotal students among freshmen-rank students, sophomore-rank students, $1xxx$-level courses, and $2xxx$-level courses. As expected, students with high betweenness centralities for $1xxx$ level courses (Figure~\ref{fig:centrality-cat1}) were taking courses corresponding to the core curriculum such as HUMA 1301, CHEM 1312, PHYS 1301, etc. When we turn our focus from $1xxx$-level courses to freshmen rank students, we find that some freshmen with high betweenness centralities were enrolled in $2xxx$ courses that most likely satisfy their general core requirements, such as GOVT 2305, GOVT 2306, ECON 2301.

\begin{figure*}[htbp]
    \centering
    
    \begin{subfigure}[b]{0.45\linewidth}
        \includegraphics[width=\linewidth]{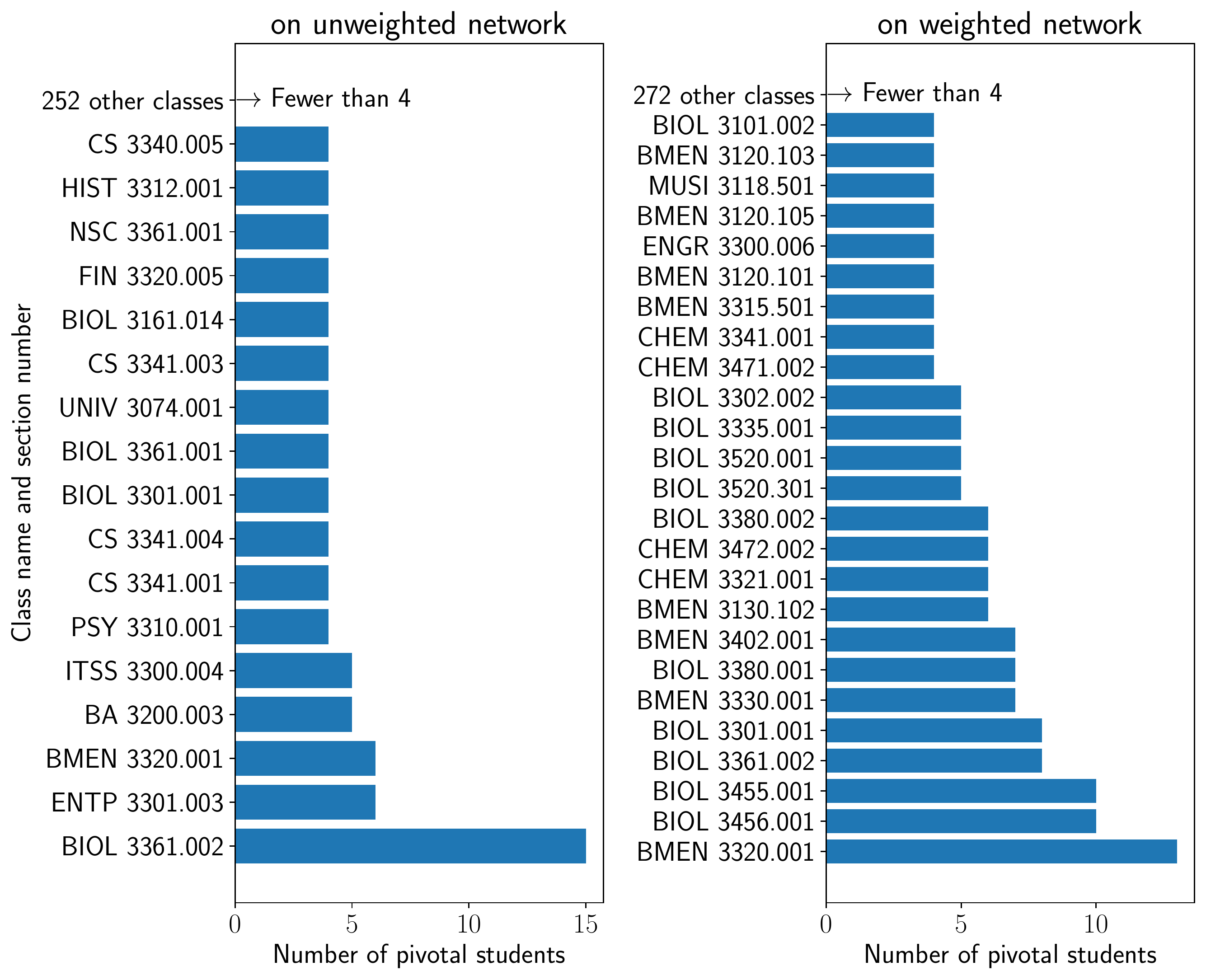}
        \caption{3000's level courses}
        \label{fig:centrality-cat3}
    \end{subfigure}
    \begin{subfigure}[b]{0.45\linewidth}
        \includegraphics[width=\linewidth]{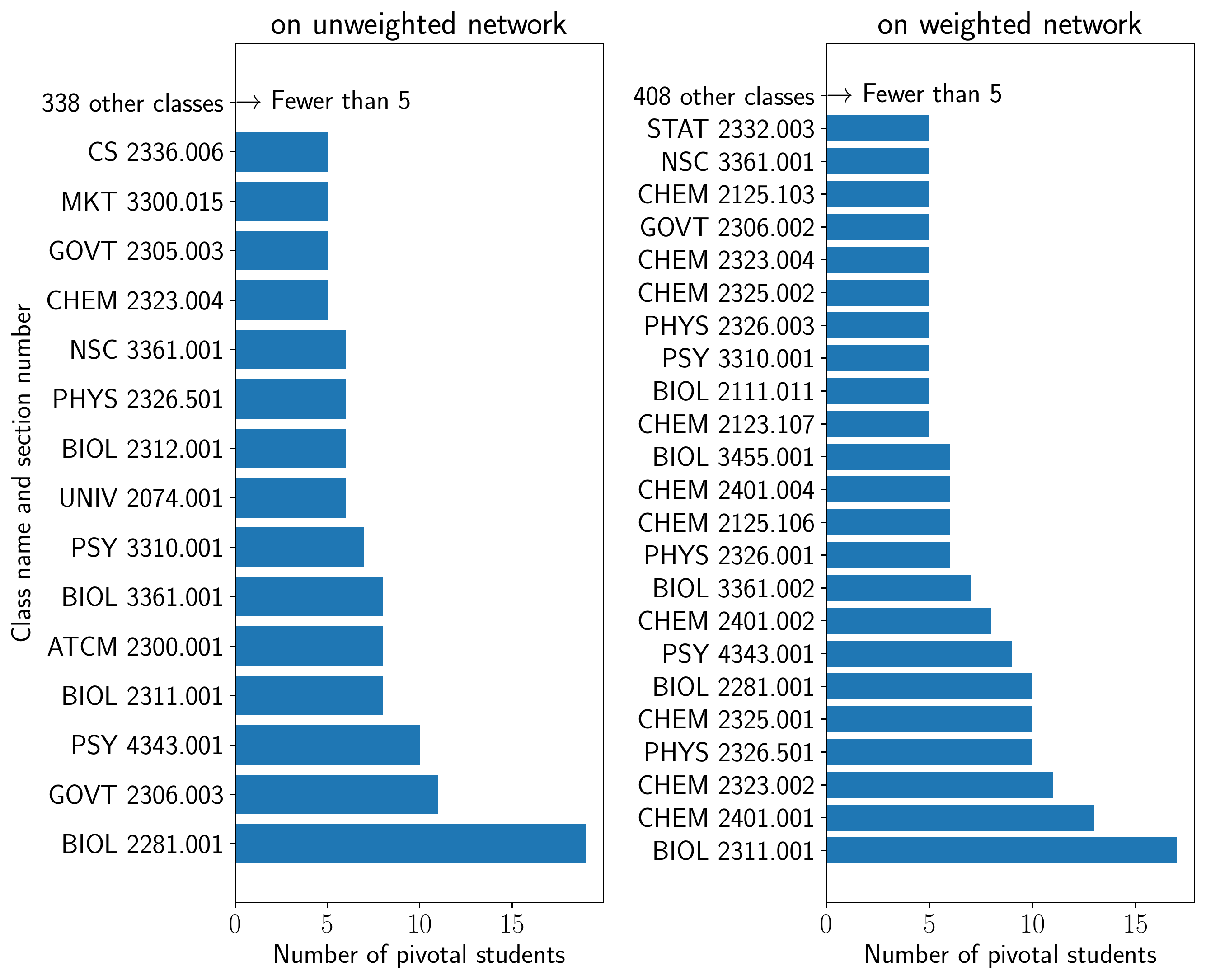}
        \caption{Junior level students only}
        \label{fig:centrality-jun}
    \end{subfigure}
    \begin{subfigure}[b]{0.45\linewidth}
        \includegraphics[width=\linewidth]{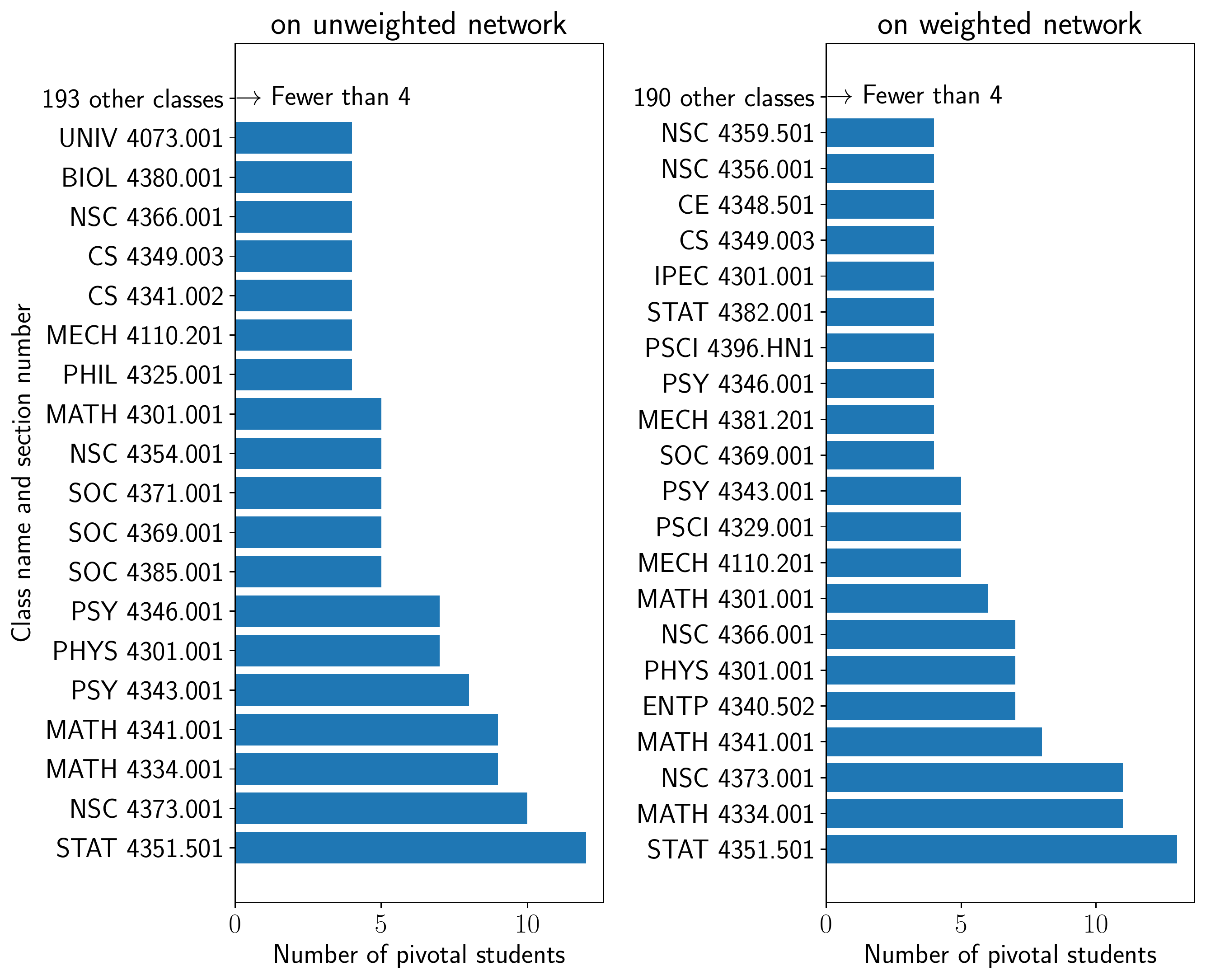}
        \caption{4000's level courses}
        \label{fig:centrality-cat4}
    \end{subfigure}
    \begin{subfigure}[b]{0.45\linewidth}
        \includegraphics[width=\linewidth]{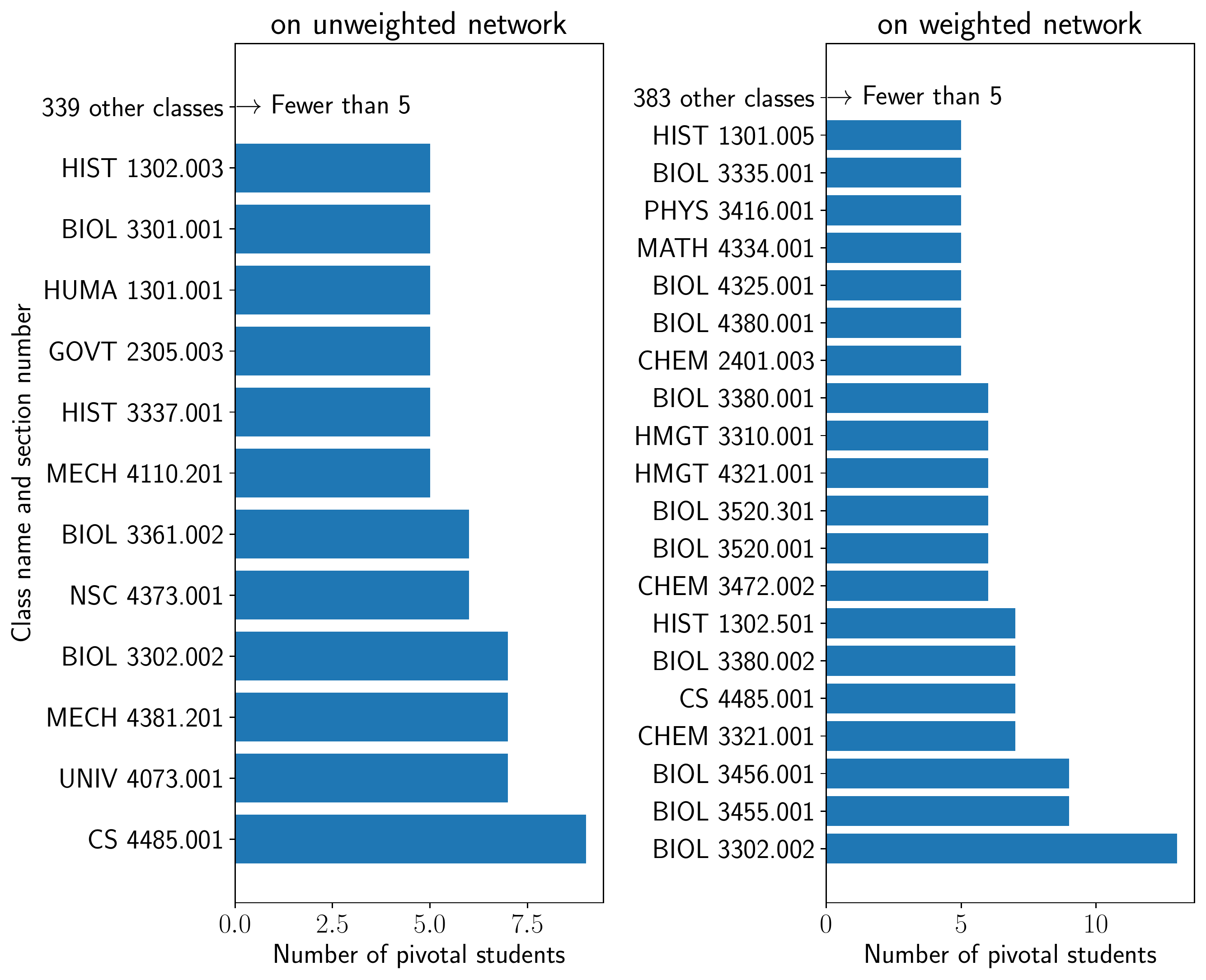}
        \caption{Senior level students only}
        \label{fig:centrality-sen}
    \end{subfigure}
    \begin{subfigure}[b]{0.45\linewidth}
        \includegraphics[width=\linewidth]{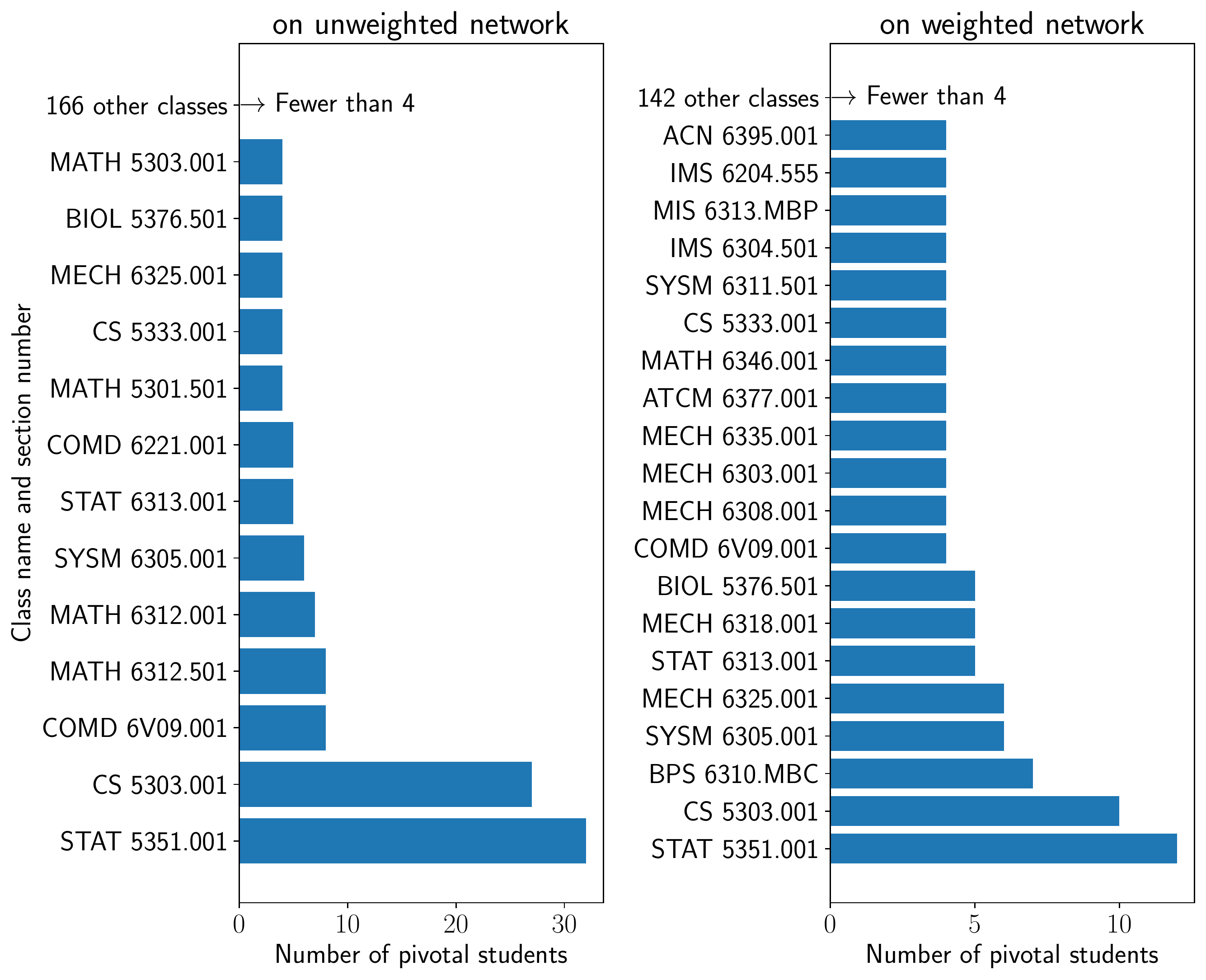}
        \caption{5000's and 6000's level courses}
        \label{fig:centrality-cat56}
    \end{subfigure}
    \begin{subfigure}[b]{0.45\linewidth}
        \includegraphics[width=\linewidth]{centralityGraduate}
        \caption{Graduate students only}
        \label{fig:centrality-cat7}
    \end{subfigure}
    \caption{The classes most frequently taken by pivotal students (nodes with high betweenness centralities) in upper division student-to-student networks. }
    \label{fig:btw-upper-division}
\end{figure*}

Figure~\ref{fig:btw-upper-division} shows the courses taken by the one hundred pivotal students when we limit the analysis to juniors, seniors, graduate students, students taking $3xxx$-level courses, students taking $4xxx$-level courses, and students taking $5xxx$ and $6xxx$ level courses. Following a course like CS 4485 across the university wide network, undergraduate student network, and senior-rank student network, an interesting trend emerges. Only six of the one hundred pivotal students in the university wide network were enrolled in CS 4485. This number increased to seven when the network was limited to undergraduate students only, and rose to eight for the network of senior-rank students. Some of these students with the highest betweenness centrality in smaller populations cease to occupy positions of similar distinction when the scope of the network is enlarged because these students are closely linked only to students within their own small group. Persuading these students to move to online learning may not have much of an impact on achieving the desired outcomes of reducing clustering and increasing the diameter of the network at the university level. On the other hand, doing the same for students who continue to be part of the pivotal student group as the scope of the network is enlarged may achieve the desired result.

\subsection{Scalpel approach: selective removal of minimal subset of courses}
\label{Scalpel}

As stated above, only when a large number of courses with high enrollment are moved online does the graph based on in-person classes show a significant increase in network diameter and average path length. If such a solution is not available to decision-makers, we explore an alternative approach that requires moving a much smaller subset of courses online. We refer to this less intrusive approach as the \textit{scalpel approach}. In this approach, we exploit the \textit{betweenness centrality} information about pivotal students, as described in Section~\ref{Courses_pivotal}.

Pivotal students lie on a disproportionately large number of short paths between pairs of students. We studied the impact of reducing the number of face-to-face classes in which such students participate, which would reduce the number of shortest paths passing through these students. Specifically, we identified all sections corresponding to the twenty-five courses with the highest enrollment of pivotal students for the unweighted network in Figure~\ref{fig:centrality-all}. \textit{We simulated moving all these sections to online instruction.} Consequently, edges between students contributed by these classes were removed from the university-wide enrollment graph. The number of sections declined from $4,956$ to $4,888$, a very modest reduction of $68$ sections (just under $1.4\%$ of all sections). Yet, the change in graph connectivity is noteworthy, as shown in Table~\ref{tab:scalpel}.

\begin{table*}[!ht]
	
	\centering \small 
	\begin{tabular}{| l | r | r |}
		\hline
		Metric                              & University & Scalpel Approach \\ \hline
		Nodes						          & 28,849      & 28,793   \\
		Edges, \textit{full graph}          & 3,714,254    & 3,302,030   \\
		Nodes, $n$                          & 27,080      & 26,954   \\
		Edges, $m$                          & 3,711,518    & 3,297,916  \\
		Average degree                      & 274.12     & 244.70698  \\
		Percent nodes in largest comp.  & 93.861     & 93.613  \\
		Average edge weight                 & 9.6231     & 9.5522 \\
		Average geodesic distance, $\ell_G$ & 2.9694     & 3.0547  \\
		Diameter of network                 & 9          & 9  \\
		Node pairs within distance $4$		& 84.219\% 	 & 82.238\% \\
		Unweighted local $C_G$              & 0.46487    & 0.47189    \\
		Unweighted global $T_G$             & 0.38386    & 0.39536 \\
		Network density, $r_G$              & 0.010122   & 0.009079  \\ \hline
	\end{tabular}
	\caption{Metrics for the largest connected component of student-to-student networks with and without the scalpel approach of removing courses.}
	\label{tab:scalpel}
\end{table*}

The scalpel approach increases average path length by approximately $0.1$, reduces the average degree of vertices significantly (by about $30$), and increases the local and global clustering coefficients while reducing network density, making it \textit{more} of a small-world network. These changes in connectivity and path length are significant considering the small number of sections that are moved online. 

By far the most remarkable impact of the scalpel approach is the reduction in the percentage of student pairs that are within a distance of $4$ of each other. As shown in Figure~\ref{fig:cumulative-dens-class-size}, removing the top $20\%$ of courses by enrollment decreased this from $84.219\%$ to  $81.477\%$, a reduction of $2.74\%$. In comparison, removing barely $1.4\%$ of classes using the scalpel approach reduced this metric by $1.98\%$ (from $84.219\%$ to  $82.238\%$) 

Hence, the scalpel approach appears to be promising. A two-pass scalpel approach could be worth considering. In the first pass, courses would be moved online based on the approach described above. In the next pass, the pivotal students for the graduate-students-only and undergraduate-students-only graphs could be identified and a limited number of courses with high number of pivotal students from corresponding populations could be moved online. This would further disrupt the small-world networks among these sub-populations.

\section{Discussion and Recommendations}
\label{Discussion}


A vaccine is unlikely to produce community immunity before the start of the fall semester of 2021. Fully online instruction is considered suboptimal for student learning and institutional finances. Hence, the key question for the remainder of Academic Year 2021, and for future occurrences of such a pandemic is this: to what extent can institutions return to face-to-face or hybrid instruction? How many classes can resume face-to-face or hybrid instruction, how large can they be, and which students can populate them?

We replicate Weeden and Cornwell's study \cite{Weeden_Cornwell_2020} by finding that students at our medium-sized public American university represent a "small-world" network susceptible to infections spread. Given the average student course load and the window for individuals with COVID-19 to move from infectious to symptomatic, a four-step threshold representing a week’s worth of courses is a good benchmark for identifying the potential for epidemic spread.

By moving as much as $60\%$ of enrollment by course section (courses with $40$ or more students) online, we were able to reduce the largest network below $75\%$ connectivity (to $64.1\%$) within four steps. In addition, by moving $80\%$ of enrollment by course section (courses with $25$ or more students) online, we were able to reduce the largest network below $50\%$ connectivity (to $33.2\%$) within four steps, and it took more than eighteen steps for $71\%$ of students to reach each other.

Building on Weeden and Cornwell, we find an inverse relationship between network connectivity and class standing (freshmen to senior), as well as a similar relationship between network connectivity and course level. Seniors are significantly less connected than freshmen are and doctoral students are significantly less connected than master’s students. Without removing any class sections from the sample, $97.5\%$ of freshmen were within four steps of one another, while $97\%$ of seniors were within four steps of one another. However, freshmen students are much more closely linked to each other than seniors: only about $24\%$ of senior pairs are within two steps of each other, whereas $91\%$ of freshmen pairs are within the same distance. In contrast, only $35.7\%$ of graduate student pairs were within four steps of one another. 

This pattern is also present in the graphs formed by considering enrollment at the various course levels. For example, $99.9\%$ of student-pairs in the $1xxx$-level graph were within four steps of each other, a number that drops to $81.9\%$ in the 4xxx-level graph. Further, it is safe to assume that this difference is not the result of low enrollment, as the $1xxx$-level graph features $9,149$ students, which is not prohibitively dissimilar to the $4xxx$-level graph’s $8,454$ students.

An even lower percentage (only $41.9\%$) of student pairs in a graph using $5xxx$ or $6xxx$ level courses (master’s level) were within four steps of one another. The percentage further declined to  $10.2\%$ of student pairs enrolled in $7xxx$ level courses (doctoral level) being within four steps of one another.

A key mechanism underpinning our findings is specialization. As students at American universities increase in class standing, they move from generalized ("core curriculum") to specialized ("major") course loads. This increases clustering within academic divisions but decreases clustering between divisions. As a result, networks of students with higher class standing tend to be split into smaller groups by major and overall connectivity decreases. Similarly, due to core curriculum and course requirements, students are more likely to take lower-level courses outside their division.

In addition, we found that certain courses have a greater potential for epidemic spread due to their centrality to the core curriculum or to majors in our largest divisions (ECS and SOM). These courses are more likely to enroll highly connected students, so taking them online may be sufficient to reduce the potential for epidemic spread without broadly restricting courses by size or division. For example, we found that $25$ courses have more than three of the $100$ most connected students in the university-wide network. Removing them from the network reduces connectivity within four steps from $84.219\%$ to $82.238\%$.

Given that this analysis is concerned with a topic of immediate relevance to thousands of higher education institutions, there are a number of questions that must be addressed when considering whether and how to convert the trends that we have observed into policy decisions. First, how generalizable are our findings to other institutions? As a medium-sized, public, four-year institution that transitions students from general to specialized courses, we expect our analysis to be most relevant to similar institutions. One potential difference is our mix of majors; in Fall 2019, $30\%$ of our undergraduate students were majoring in the School of Engineering and Computer Science while a further $22\%$ were majoring in the School of Management. Beyond the core curriculum, programs from both schools (e.g., Computer Science, Finance) require courses in mathematics, natural sciences (e.g., physics), and/or social sciences (e.g., economics, political science). Relative to an institution with a larger share of arts and humanities or social sciences students, students at our institution may be more connected due to our programmatic distribution.

Furthermore, as Weeden and Cornwell note, our analysis is necessarily limited to academic networks. It does not consider residential patterns, social activities, contact in research spaces (libraries or laboratories), transit between courses, shared instructors or class spaces, transit modes to and from the institution (mass vs. personal transit), or exposure to the broader community through activities such as grocery shopping. Additionally, our analysis does not consider how the presence or absence of mitigating policies, such as testing and contact tracing, mandated masking, or restrictions on high-transmission activities (such as indoor dining), affect spread at an institution. The impact of each factor will vary by student demographics; the characteristics of the surrounding community; and institutional, local, and state policies.

In general, we can identify factors that are likely to reinforce the difference in the susceptibility to infectious spread and other factors likely to reduce this difference. As lower-division students are more likely to live in dorms, they are in closer contact with other students through shared bedrooms, bathrooms, and common areas. Similarly, undergraduates are more likely than graduate students to engage in extracurricular activities that increase the risk of transmission, notably fraternities/sororities and campus athletics. Conversely, upper division students and graduate students are more likely to have legal access to bars, which are associated with a high risk of transmission, or to live with non-college students (e.g., roommates or a spouse), which increases the risk of spread from outside the student community. In addition, we should not discount that older students are generally less susceptible to risk-taking behaviors \cite{arain2013maturation}.  Given the relatively low number of freshmen who live on campus and the relatively high share of transfer and older ($25+$) undergraduates, UT Dallas undergraduates are likely less connected outside the classroom than students at the archetypical flagship or major private university.

A common strategy for Fall 2020 was to focus on bringing lower-division students back to campus in order to maintain the "freshman experience". There are academic and financial incentives for pursuing this strategy. Student attrition is highest from the first to the second academic year, making that early experience crucial for habituating students to the rigors of higher education and strengthening connections to their institution (and each other). Financially, first year students are the most likely to utilize on campus, university-owned dormitories, which represent a significant, fixed cost for institutions.

Our analysis indicates that a strategy returning lower-division students to campus runs contrary to the goal of minimizing infectious spread through an academic network in the absence additional steps to reduce connectivity. One potential mitigating step would be to recreate the specialization of upper-division students through academic \textit{pods}, restricting face-to-face or hybrid instruction to courses in their academic division (Engineering, Management, etc.) while limiting student enrollment in courses outside their division to online instruction. This method would maximize student habituation to their discipline, where a student will spend most of their academic careers, while minimizing the risk of infectious spread across divisions. Furthermore, though housing arrangements for Academic Year 2021 are effectively locked in, institutions facing similar challenges in the future could combine academic and habitation choices, assigning students to dormitories by academic division. This strategy of creating \textit{bubbles}, combined with a rigorous testing regime, proved successful for major American sports leagues in 2020. However, our analysis also indicates that institutions with sufficient analytic capacity might also take a \textit{scalpel} approach to instruction modes by identifying the specific courses most likely to facilitate epidemic spread.

\section{Conclusion}
\label{Conclusion}

In this work we have performed a graph-theoretic analysis of interconnections among students in a medium-sized public university. This analysis is similar to that performed by Weeden and Cornwell \cite{Weeden_Cornwell_2020} for a medium-sized private university. Students, through enrollment in various courses, form a small world network. 

Due to the comparatively larger section sizes in the public university considered in this study, the solution proposed by Weeden and Cornwell to prevent the spread of contagion through classroom contact may not be as effective. This work proposes a complementary solution, the \textit{scalpel} approach, that disrupts paths for the spread of contagion by moving courses that contribute to high \textit{betweenness centrality} to online instruction. Another recommendation is to maintain classroom teaching of highly specialized graduate and some upper-division courses as students in these courses tend to be tightly knit with little to know connection with other students, while moving all other courses online.

While the development of vaccines may result in some restoration of normalcy by Fall 2021, recommendations provided in this work could be useful for policy makers if faced with similar pandemics in the future.


\bibliography{small_world_paper}

\end{document}